\documentclass[useAMS,fleqn, usenatbib, final]{mnras}

\usepackage{lipsum}
\usepackage{multirow}
\usepackage{import}
\usepackage{tabularx}
\usepackage{glossaries}
\usepackage[titletoc,title]{appendix}

\usepackage{graphicx}

\usepackage{amsmath}
\usepackage{amssymb}
\usepackage{amsfonts}
\usepackage{mathtools}
\usepackage{bm}

\usepackage[utf8]{inputenc}

\usepackage[table]{xcolor}
\usepackage{url}

\usepackage{rotating}
\usepackage{booktabs}

\usepackage{dcolumn}
\newcolumntype{d}{D{.}{.}{-1}}
\newcolumntype{b}[1]{D{(}{(}{#1}}

\definecolor{darkgreen}{rgb}{0.0, 0.5, 0.0}

\usepackage{xspace}
\usepackage[normalem]{ulem}

\newcommand{\num}{1}
\newcommand{\portsmouth}{Institute of Cosmology and Gravitation, University of Portsmouth, Portsmouth, PO1 3FX, UK}
\newcommand{\model}{\mathfrak{M}}

\newacronym{LIGO}{LIGO}{Laser Interferometer Gravitational-Wave Observatory}
\newacronym{LVK}{LVK}{LIGO-Virgo-KAGRA}
\newacronym{NR}{NR}{numerical relativity}
\newacronym{GR}{GR}{General Relativity}
\newacronym{GW}{GW}{gravitational-wave}
\newacronym{PSD}{PSD}{Power Spectral Density}
\newacronym{MCMC}{MCMC}{Markov Chain Monte Carlo}
\newacronym{SNR}{SNR}{signal-to-noise ratio}
\newacronym{IMBH}{IMBH}{Intermediate Mass Black Hole}
\newacronym{SMBH}{SMBH}{Supermassive Black Hole}
\newacronym{ISCO}{ISCO}{Innermost Stable Circular Orbit}
\newacronym{QNM}{QNM}{quasinormal mode}
\newacronym{HPC}{HPC}{High Performance Compute}

\title[The Missing Multipole Problem]{The Missing Multipole Problem: Investigating biases from model starting frequency in gravitational-wave analyses}

\author[Ursell \emph{et al.}]{
\parbox{\textwidth}{
R.~Ursell$^{\num}$
,
C.~Hoy$^{\num}$\thanks{charlie.hoy@port.ac.uk}
,
I.~Harry$^{\num}$
,
L.~K.~Nuttall$^{\num}$
}
\vspace{0.2cm}\\
$^1$\portsmouth \\
}
\date{Accepted XX. Received YY; in original form ZZ}
\pubyear{2025}

\begin{document}
\label{firstpage}
\pagerange{\pageref{firstpage}--\pageref{lastpage}}

\maketitle

\begin{abstract}
Our ability to infer the true source properties of colliding black holes from gravitational wave observations requires not only accurate waveform models but also their correct use. A key property when evaluating time-domain models is when to start the waveform: choosing a time that is too late can omit low-frequency power from higher order multipoles. By focusing on binary systems with total mass $\ge 200 \, M_{\odot}$, we show that current detectors are sensitive to this missing power and biased source properties can be obtained. We show that for systems with total mass $\lesssim 300 \, M_{\odot}$, mass ratio $\gtrsim 0.33$, and signal-to-noise ratio $\rho \gtrsim 20$, templates starting at $20 \, \mathrm{Hz}$ recover biased source properties. As the total mass increases, and the component masses become more asymmetric, templates starting from $13 \, \mathrm{Hz}$ recover biased properties. If the gravitational-wave signal is observed at signal-to-noise ratio $\rho < 20$, time-domain models can start from $20\, \mathrm{Hz}$ as statistical uncertainties dominate. 

\end{abstract}

\section{Introduction} \label{sec:intro}
Fast and accurate \gls{GW} models are essential for extracting astrophysical information from observed signals through Bayesian inference. This method relies on models to produce millions of possible theoretical signals -- referred to as ``templates'' -- to match against the data~\citep{LIGOScientific:2025yae}. Typically, these analyses are performed in the frequency-domain between a minimum $(f_{\mathrm{low}})$ and maximum $(f_{\mathrm{max}})$ frequency~\citep[although see][for time-domain implementations]{Carullo:2019flw,Isi:2021iql,Miller:2023ncs}.

\Gls{GW} signals can be decomposed into a sum of -2 spin-weighted spherical harmonics,
with the quadrupole $(\ell=2, m=\pm2)$ being the lowest-order and typically the most dominant contribution~\citep{Goldberg:1966uu,Thorne:1980ru}. The amplitudes of additional higher-order multipole moments vary in parameter space, with many becoming more significant in systems with asymmetric mass ratios, inclined orbital planes, and high binary masses~\citep{Mills:2020thr, Khan:2019kot, CalderonBustillo:2015lrt}. For most sources observed with ground-based \gls{GW} detectors~\citep{LIGOScientific:2014pky,VIRGO:2014yos,KAGRA:2020tym}, the $(\ell, |m|) = (3, 3)$ multipole is the most significant higher order correction~\citep{Mills:2020thr}; the $(3, 3)$ multipole was first observed in GW190412 and GW190814~\citep{LIGOScientific:2020stg,LIGOScientific:2020zkf,Hoy:2024wkc}.

Models can be defined in the time or frequency-domain. When time-domain models are converted to the frequency domain for analyses, a non-negligible fraction of each higher order multipole is excluded from the signal; if the $(\ell, |m|) = (2, 2)$ multipole starts at $20\, \mathrm{Hz}$, we will miss power from the $(3, 3)$ multiple between $20 - 30\, \mathrm{Hz}$ and the $(4, 4)$ multiple between $20 - 40\, \mathrm{Hz}$. The absence of this low-frequency content -- referred to as the ``missing multipole problem" -- may introduce biases in our estimates for the source properties, especially for short-duration signals with limited inspiral.

\cite{Islam:2023zzj} investigated the missing multipole problem previously. They concluded that missing frequency content from higher order multipoles had only a minor impact on Bayesian analyses. However, their analysis focused on binary black hole systems with moderate total masses and near-equal component masses, where higher-order multipoles are not expected to contribute significantly to the total signal power.

In light of recent short duration \gls{GW} observations, for example GW190521~\citep{LIGOScientific:2020iuh}, GW231123\_135430~\citep{LIGOScientific:2025rsn} (hereafter GW231123) and others~\citep{LIGOScientific:2021usb, Wadekar:2024zdq, Ruiz-Rocha:2025yno}, we revisit the study in \cite{Islam:2023zzj} and perform a detailed systematic study to investigate how template starting frequency biases Bayesian analyses.
We show that for light \gls{IMBH} systems with total binary mass $M \lesssim 300 \, M_{\odot}$ (in the detector-frame), mass ratios $q \gtrsim 0.33$, and network \glspl{SNR} $20 \lesssim \rho \lesssim 70$, templates starting from $f_{\mathrm{low}}$ can produce biased source properties; see Appendix~\ref{sec:appendix_notation} for a description of the notation used in this paper. For systems with total mass $\gtrsim 300 \, M_{\odot}$, templates starting from $2/3 \, f_{\mathrm{low}}$ recover biased properties. However, as the signal-to-noise ratio of the observed signal decreases $(< 20)$, statistical uncertainties dominate and missing low-frequency \gls{GW} power can be ignored. We finally show that our recommendations apply for a real \gls{GW} signal observed in the fourth \gls{GW} observing run.

This paper is structured as follows: in Secs.~\ref{sec:models} and \ref{sec:pe} we give a brief overview of the waveform models and analysis methods used. In Sec.~\ref{sec:results}, we present a detailed comparison of parameter estimation results with and without low-frequency higher order multipole content. In Sec.~\ref{sec:GW231123} we show that our recommendations remain consistent with real \gls{GW} signals. Finally, in Sec.~\ref{sec:discussion}, we conclude with a discussion of the implications for \gls{GW} observations.

\section{Gravitational Wave Models} \label{sec:models}

\begin{figure*}
    \includegraphics[width=0.88\textwidth]{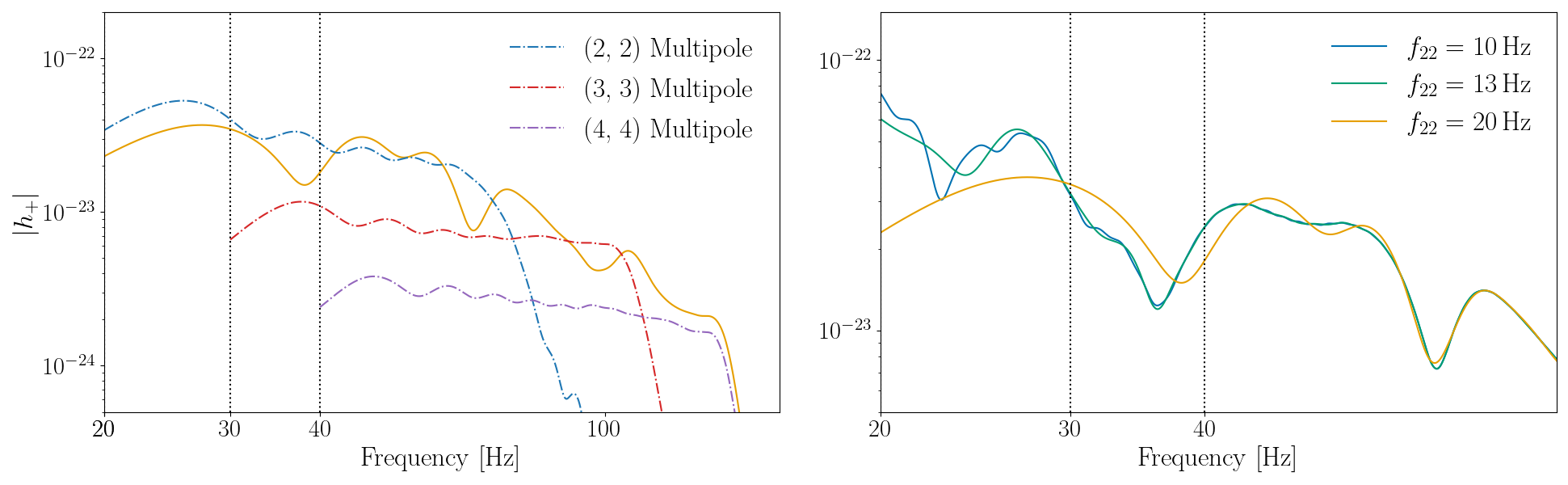}
    \caption{Plot showing the amplitude of the plus polarization 
    $h_{+}$ of a Fourier transformed \gls{GW} signal for a simulated light \gls{IMBH} system produced with the {\textsc{NRSur7dq4}} waveform model~\citep{Varma:2019csw}. The left panel shows a \gls{GW} signal when the starting frequency of the $(\ell, m) = (2, 2)$ multipole is $f_{22} = 20\, \mathrm{Hz}$. We also show a selection of higher order multipoles for the $20 \, \mathrm{Hz}$ case. The right panel compares the same \gls{GW} signal produced with starting frequencies $f_{22} = 10 \, \mathrm{Hz}$ (blue), $f_{22} = 13 \, \mathrm{Hz}$ (green), and $f_{22} = 20 \, \mathrm{Hz}$ (orange) over a reduced frequency range. In both panels, the black dotted lines indicate the starting frequency of the $(\ell, m) = (2, 2), (3, 3)$ and $(4, 4)$ multipoles, see Eq.~\ref{equ:Harmonic_multipoles}. In all cases, the gravitational-wave is produced from the same simulated \gls{IMBH} system with total mass $M = 300\, M_{\odot}$, mass ratio $q=0.25$, spin magnitudes $\chi_{1} = \chi_{2} = 0.7$ and observed an inclination angle angle of $\theta_{\mathrm{JN}} = \pi / 3\, \mathrm{rad}$.}
    \label{fig:waveform}
\end{figure*}

Numerous models are now available for analysing \gls{GW} data~\citep[see e.g.][]{Varma:2019csw,Estelles:2020twz,Pratten:2020ceb,Thompson:2023ase,Ramos-Buades:2023ehm,Colleoni:2024knd,Hamilton:2025xru,Estelles:2025zah}.
These models tend to combine perturbative methods, numerical solutions, and qualitative insights to produce fast and accurate theoretical signals.
However, time and computational limitations mean that these models only include a subset of higher order multipole moments -- up to $\ell = \ell_{\mathrm{max}}$ and $m = m_{\mathrm{max}}$ -- see Appendix~\ref{sec:appendix_hms} for details.
Among those currently available, {\textsc{NRSur7dq4}} is on average the most accurate for \gls{GW} analyses~\citep{Varma:2019csw,Islam:2023zzj}.
{\textsc{NRSur7dq4}} models a \gls{GW} signal as a sum of -2 spin-weighted spherical harmonics up to $\ell_{\mathrm{max}} = 4$, and includes the effects of spin precession~\citep{Apostolatos:1994mx}.

\gls{GW} models can be defined in the time or frequency-domain, with both offering their advantages and disadvantages~\citep{LIGOScientific:2025yae}. For models defined in the time-domain, each spherical harmonic starts at a time $t=t_{0}$. At large separations (during the early inspiral), the time-domain oscillation frequency of each $(\ell, m)$ harmonic is $f = m\Omega$, where $\Omega$ is the binary orbital frequency~\citep{Kidder:2007rt}. Under this regime, if the initial frequency of the dominant quadrupole at time $t = t_0$ is $f(t_{0})=f_{22}$, the initial frequency of each subsequent multipole is~\citep{Buonanno:2002fy, London:2017bcn},
\begin{equation} \label{equ:Harmonic_multipoles}
    f_{\ell m} (t_{0}) = \frac{m}{2}f_{22} \, .
\end{equation}
At closer separations (during late inspiral and merger), the frequency of each $(\ell, m)$ multipole scales approximately as $f = \ell\, \Omega$~\citep{Leaver:1985ax,Berti:2005ys}. Since \gls{GW} analyses are typically performed in the frequency-domain~\citep{LIGOScientific:2025yae}, this poses a challenge for time-domain models as each harmonic starts at different characteristic frequencies $f_{\ell m}(t_{0})$. This feature does not impact models defined in frequency-domain, as the initial frequency of each multipole can be set to the same value, \emph{i.e.} $f_{\ell m} \equiv f_{22}$. 

Ground-based \gls{GW} detectors are most sensitive between $\sim 20$--$1000 \, \mathrm{Hz}$~\citep{LIGOScientific:2014pky,VIRGO:2014yos,KAGRA:2020tym,aLIGO:2020wna} and as such, analyses typically start from $f_{\mathrm{low}} = 20\, \mathrm{Hz}$. For time-domain models to accurately describe higher order multipole content from $f = f_{\mathrm{low}}$, the template must be started such that $f_{22} = f_{\mathrm{low}} / m_{\mathrm{max}}$. If the model is started at later times/frequencies
higher-order multipoles will be partially or entirely excluded from the waveform. This effect is particularly pronounced for short-duration \gls{GW} signals, where there is insufficient time for these multipoles to enter the analysis band before merger.

In Fig.~\ref{fig:waveform} we show a theoretical time-domain GW signal in the frequency domain. This signal was produced by {\textsc{NRSur7dq4}} for a binary black hole system with component masses $m_1 = 240 \, M_{\odot}, \ m_2 = 60 \, M_{\odot}$ and spin magnitudes $\chi_{1} = \chi_{2} = 0.7$. We see that when the dominant quadrupole starts at $f_{22} = 20 \, \mathrm{Hz}$, the $(3,3)$ and $(4,4)$ multipoles do not contribute until approximately $30 \, \mathrm{Hz}$ and $40 \, \mathrm{Hz}$, respectively\footnote{There will be some contribution below $30 \, \mathrm{Hz}$ and $40 \, \mathrm{Hz}$ from the $(3,3)$ and $(4,4)$ multipoles respectively due to spectral leakage from the Fourier transform. In this plot, we have removed this by placing a frequency mask below the expected starting frequencies.}. We see that this leads to visible differences between theoretical signals produced with different starting frequencies. 
Given that {\textsc{NRSur7dq4}} includes spherical harmonics up to $\ell_{\mathrm{max}} = m_{\mathrm{max}} = 4$, a starting frequency of $f_{22} = 10\, \mathrm{Hz}$ is needed such that all GW content is accurately captured from $f_{\mathrm{low}} = 20\, \mathrm{Hz}$.

\section{Gravitational Wave Bayesian Inference} \label{sec:pe}

Bayesian inference is a statistical tool for estimating astrophysical properties given some observational data $d$, and a model for the astrophysical phenomenon $\model$.
These properties are described by the \emph{posterior probability distribution}, $P(\theta \mid d, \model)$, which quantifies the probability of the parameters $\theta$ given the data and model. Bayes’ theorem defines the posterior probability distribution as:
\begin{equation}
\label{equ:Bayes Theorem}
P(\theta \mid d, \model) = \frac{\mathcal{L}(d \mid \theta, \model) \, \Pi(\theta \mid \model)}{\mathcal{Z}},
\end{equation}
where $\mathcal{L}(d \mid \theta, \model)$ is the likelihood, quantifying how well a model evaluated at a set of parameters explains the observed data; $\Pi(\theta \mid \model)$ is the prior, encoding our knowledge or assumptions about the astrophysical parameters before observing the data; and $\mathcal{Z} = \int \mathcal{L}(d \mid \theta, \model) \, \Pi(\theta \mid \model)\, d\theta$ is the evidence, which normalises the posterior.

When multiple Bayesian analyses are performed with e.g. different astrophysical models, comparing evidences is useful as it quantifies how much the data supports one model over another. Often referred to as the Bayes factor, it is computed through:
\begin{equation}
    \label{equ:bayes_factor}
    \mathcal{B} = \frac{\mathcal{Z}_{1}}{\mathcal{Z}_{2}}.
\end{equation}
A Bayes factor $\mathcal{B} > 1$ indicates that the data favours $\model_{1}$, while $\mathcal{B} < 1$ suggests a preference for $\model_{2}$. According to the widely used scale from~\cite{Kass:1995loi}, $\log_{10} \, \mathcal{B} > 2$ constitutes decisive evidence in favour of $\model_{1}$, $1 < \log_{10} \, \mathcal{B} < 2$ implies strong evidence in favour of $\model_{1}$ and $0.5 < \log_{10} \, \mathcal{B} < 1$ indicates substantial evidence in favour of $\model_{1}$.

For \gls{GW} astronomy the likelihood is well known~\citep[see e.g.][]{Veitch:2014wba,Thrane:2018qnx}.
For a single detector, the (log) likelihood is simply\footnote{The likelihood also includes an additional term describing the noise covariance. Under the assumption that the noise is Gaussian and stationary, the noise covariance matrix is the identity matrix and often excluded for simplicity. When this assumption is no longer valid, the noise covariance should be included, see e.g.~\cite{Edy:2021par} for details.},
\begin{equation}
    \label{equ:Log_Likelihood}
    \ln \mathcal{L} \propto {\langle d - \model(\theta)\mid d - \model(\theta) \rangle},
\end{equation}
where $\langle a \mid b \rangle$ is the inner product~\citep{Finn:1992wt,Owen:1995tm},
\begin{equation}
    \label{equ:Inner_Product_simple}
    \langle a\mid b \rangle \propto \int_{f_{\mathrm{low}}}^{f_{\mathrm{max}}} \mathrm{d}f \, \frac{a(f)b^*(f)}{S_n(f)} \, ,
\end{equation}
$S_n(f)$ is the \gls{PSD}, $f_{\mathrm{min}}$ ($f_{\mathrm{max}}$) are the minimum (maximum) frequencies considered, and $*$ represents the complex conjugate. For a network of $N$ \gls{GW} detectors, Eq.~\ref{equ:Log_Likelihood} becomes,
\begin{equation}
    \ln \mathcal{L} \propto \sum_{i = 0}^{N}{\langle d_{i} - \model(\theta)\mid d_{i} - \model(\theta) \rangle}.
\end{equation}

Assuming a quasi-circular binary black hole merger (\emph{i.e.} zero eccentricity of the orbital plane), $\theta$ is a 15 dimensional vector: 8 intrinsic dimensions describing the individual component masses and spin angular momenta of each black hole, and 7 extrinsic dimensions describing the binary's inclination angle, orbital phase, luminosity distance, right ascension, declination, polarisation angle, and coalescence time. See Appendix~\ref{sec:appendix_notation} for details.

To assess the performance of a \gls{GW} model, Bayesian analyses are often performed on simulated \gls{GW} data that contain theoretical signals of known parameters $\theta_{\mathrm{inj}}$ and noise $n(t)$. The true parameters of the signal are compared with the inferred posterior distribution to determine the accuracy with which the model describes the signal. The theoretical signal, $h(\theta_{\mathrm{inj}})$, can be injected into real or synthetic \gls{GW} detector noise, or injected into ``zero-noise'' where $n(t) = 0 \ \forall t$. In the zero-noise approximation, the inferred posterior distribution will peak at the true parameters $\theta_{\mathrm{inj}}$ when the template perfectly describes the theoretical signal, \emph{i.e.} $\model(\theta_{\mathrm{inj}}) = h(\theta_{\mathrm{inj}})$, and uniform priors are employed in all dimensions. When the template does not perfectly describe the theoretical signal, the posterior will be biased.
A posterior distribution is often said to be biased when
the marginalised one-dimensional distribution does not contain the true value within the 90\% credible interval. Alternative metrics, such as the root-mean-square deviation of the posterior from the injected value~\citep{Knee:2021noc}, have been previously used to quantify the bias in posterior distributions~\citep[see e.g.][]{Akcay:2025rve}.

In this work, we use the Mahalanobis recovery score, $r_{\mathrm{M}}$. This score quantifies biases in posterior distributions by determining the fraction of posterior realisations whose $90\%$ credible intervals contain the true value. By construction, $r_{\mathrm{M}}$ ranges between $0$ (no posterior realisations capture the truth) and an upper bound set by the chosen credible interval ($\sim 0.9$ for a $90\%$ interval under Gaussian assumptions). A higher score indicates better recovery of the injected value, while lower scores highlight systematic biases. Importantly, the score does not test whether the posterior mean itself captures the true value within its $90\%$ credible interval; instead, it measures the proportion of posterior draws that would.

To compare results from different analyses, we define the Mahalanobis difference, $\Delta r_{M}$, as the difference between their recovery scores. We normalize the $\Delta r_{M}$ to lie between $-1$ (analysis 2 outperforms analysis 1) and $+1$ (analysis 1 outperforms analysis 2). A value of zero indicates comparable performance. Full details are provided in Appendix~\ref{sec:appendix_bias}.

\section{Results} \label{sec:results}

Although numerous stochastic sampling methods are available, such as \gls{MCMC}~\citep{metropolis1949monte}, nested sampling~\citep{Skilling:2004pqw, Skilling:2006gxv}, an others~\citep{Lange:2018pyp, Tiwari:2023mzf}, in this work we perform Bayesian inference via the nested sampling algorithm \textsc{dynesty}~\citep{Speagle:2019ivv}, as implemented in  \textsc{Bilby}~\citep{Ashton:2018jfp,Romero-Shaw:2020owr}.
We evaluate the likelihood (see Equation~\ref{equ:Log_Likelihood}) over the frequency range $f_{\mathrm{low}} = 20 \, \mathrm{Hz}$ to $f_{\mathrm{max}} = 2048 \, \mathrm{Hz}$. We assume a network of two Advanced LIGO detectors~\citep{LIGOScientific:2014pky} and one Advanced Virgo detector~\citep{VIRGO:2014yos} operating at their design sensitivity for the fourth \gls{GW} observing run~\citep{O4PSD}. All other settings, including the number of live points ($n_{\mathrm{live}} = 1000$) and the uninformative prior distributions used, are the same as previous \gls{LVK} papers, see e.g.~\citep{KAGRA:2021vkt, LIGOScientific:2025yae}. All simulated signals, $h(\theta_{\mathrm{inj}})$, and template waveforms, $\model(\theta)$, are generated with {\textsc{NRSur7dq4}}~\citep{Varma:2019csw} at a reference frequency $20\, \mathrm{Hz}$.

Owing to the limited length of {\textsc{NRSur7dq4}}, we place additional cuts on the prior volume: we restrict the mass ratio $q = m_{2} / m_{1} > 0.2$ and total mass $M = m_{1} + m_2 > 150\, M_{\odot}$. We consistently used $f_{22} = 10\, \mathrm{Hz}$ for all injected signals. The starting frequency of the template was varied for each injection in 3 separate analyses. We chose starting frequencies $f_{22} = 10\, \mathrm{Hz}, 13\, \mathrm{Hz}$, and $20\, \mathrm{Hz}$. These values were chosen to progressively control the visibility of higher-order multipoles within the analysis band; see Eq.~\ref{equ:Harmonic_multipoles}.

For the lowest-frequency template, $f_{22} = 10 \, \mathrm{Hz}$, all multipoles up to $\ell_{\mathrm{max}} = m_{\mathrm{max}} = 4$ enter the band before the likelihood evaluation begins at $20 \, \mathrm{Hz}$. This represents our baseline ``unbiased” scenario. When $f_{22} = 13 \, \mathrm{Hz}$, only multipoles up to $\ell_{\mathrm{max}} = m_{\mathrm{max}} = 3$ are accessible when the analysis starts. At the largest frequency considered in this study, $f_{22} = 20 \, \mathrm{Hz}$, only the dominant $(2,2)$ multipole is present at $20 \, \mathrm{Hz}$; the \emph{e.g.} $(3,3)$ and $(4,4)$ multipoles remain inaccessible until approximately $30 \, \mathrm{Hz}$ and $40 \, \mathrm{Hz}$, respectively.

In this study, we focus on light \gls{IMBH} sources. While definitive bounds have not been formally agreed, light \glspl{IMBH} binaries observable with current \gls{GW} detectors are generally considered to have component masses between $100 - 250 M_{\odot}$. If light \gls{IMBH} sources are a product of second-generation mergers, \emph{i.e.} each black hole has been produced from a previous merger, both components are likely rapidly spinning, $\gtrsim 0.7$~\citep{Rezzolla:2007xa,Buonanno:2007sv,Healy:2014yta,Hofmann:2016yih}. Other binary properties remain uncertain~\citep[see e.g.][]{Doctor:2019ruh}.

\subsection{The Golden Case} \label{sec:golden}
First, we inject a \gls{GW} signal produced from a binary with total mass $M=300\, M_{\odot}$, mass ratio $q=0.25$ and inclination angle, defined as the angle between the line of sight and total angular momentum, $\theta_{\mathrm{JN}} = \pi / 3\, \mathrm{rad}$. The spin magnitudes for each black hole were $0.7$ and the spin tilt angles were $0.2$ and $0.8$ respectively. The luminosity distance was chosen such that the network \gls{SNR} was $\rho=75$. Although the SNR is
high when compared to the majority of observations~\citep{LIGOScientific:2025slb}, it remains
less than the highest \gls{SNR} observation to date~\citep{KAGRA:2025oiz} and comparable to the expected SNR at which GW150914~\citep{LIGOScientific:2016aoc}
would have been observed with current detector sensitivities~\citep{Gaebel:2017zys}. All other parameters were randomly chosen. 

\begin{figure}
    \includegraphics[width=0.4\textwidth]{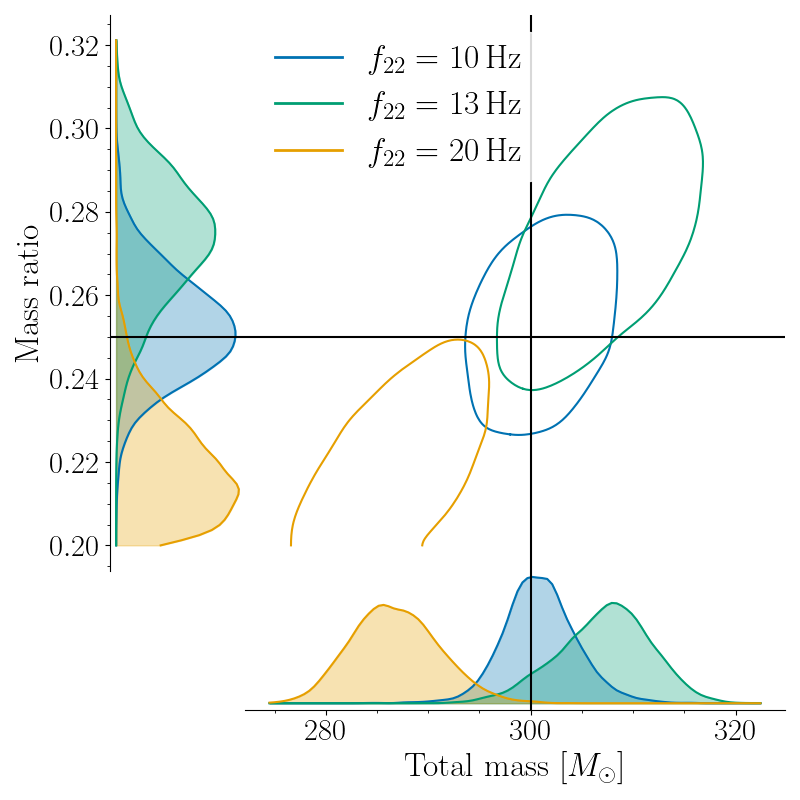}
    \caption{The two-dimensional marginalised posterior distribution for the inferred total mass $M$ and mass ratio $q$ for our golden injection, see Sec.~\ref{sec:golden}. In blue, green and orange we show the posterior distribution obtained when the template starting frequency is $f_{22} = 10, 13$ and $20\, \mathrm{Hz}$ respectively. The contours represent the inferred 90\% credible interval and the black horizontal and vertical lines show the true value.}
    \label{fig:golden}
\end{figure}

Fig.~\ref{fig:golden} shows that the $f_{22} = 10 \, \mathrm{Hz}$ analysis closely recovers the injected total mass and mass ratio of the binary, while the $f_{22} = 13 \, \mathrm{Hz}$ and $f_{22} = 20 \, \mathrm{Hz}$ cases exhibit increasing bias. Specifically, the $f_{22} = 13 \, \mathrm{Hz}$ ($f_{22} = 20 \, \mathrm{Hz}$) case overestimates (underestimates) the \gls{IMBH} masses and underestimates (overestimates) the mass ratio. Notably, the $f_{22} = 20 \, \mathrm{Hz}$ case fails to capture the injected value within the $90 \%$ credible interval.
This behaviour is reflected in the Mahalanobis recovery scores, with $r_{M} = 0.87$, 
$0.52$,
and $0.11$ for the $f_{22} = 10$, $13$, and $20 \, \mathrm{Hz}$ analyses respectively.
The Mahalanobis recovery scores suggest that for this configuration, missing power from the $(3,3)$ multipole leads to a greater loss of accuracy than excluding the $(4,4)$. This is also reflected by the Bayes factors.

The recovery of the other binary parameters showed broadly consistent trends: the two-dimensional marginalised posterior distributions for the recovered spin magnitudes as well as effective parallel spin and mass ratio, show a clear separation in performance between the three starting frequencies.
Interestingly, we see that the $f_{22} = 13 \, \mathrm{Hz}$ case recovers a posterior distribution comparable to the $f_{22} = 10 \, \mathrm{Hz}$ case for the luminosity distance and inclination angle.

This result is not specific to 15 dimensions. It is possible to achieve biased posterior distributions for templates with $f_{22} = 20\, \mathrm{Hz}$ even when considering a simple two-dimensional example. In Appendix.~\ref{sec:toy} we consider an example where the only parameters in the template are the binary component masses and show that biased posterior distributions can be obtained.

\begin{figure*}
    \includegraphics[width=0.92\textwidth]{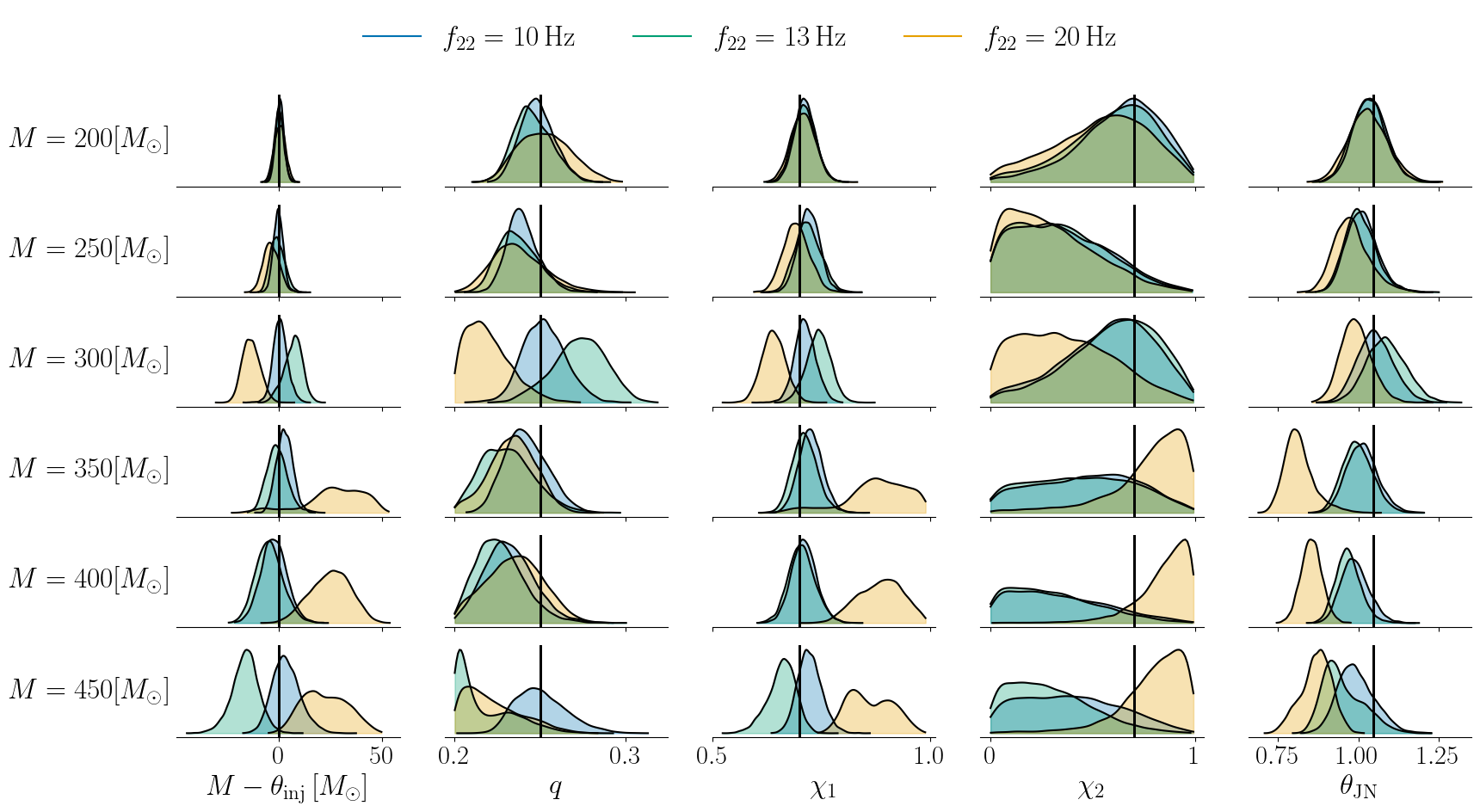}
    \caption{The one-dimensional marginalised posterior distribution for the inferred total mass $M$, mass ratio $q$, primary and secondary spin magnitudes $\chi_{1}$, $\chi_{2}$ respectively and inclination angle of the binary $\theta_{\mathrm{JN}}$ when varying the total mass of the binary. In blue, green and orange we show the posterior distribution obtained when the template starting frequency is $f_{22} = 10, 13$ and $20\, \mathrm{Hz}$ respectively. In all panels, the black solid vertical line shows the true source properties of the binary. For the left column we shift the posterior distribution by the injected total mass to centre the true value around $0\, M_{\odot}$. The different rows shows the results for different total mass injections.}
    \label{fig:total_mass_posteriors}
\end{figure*}

\subsection{Detailed injection/recovery analysis}
Now we discuss the results from our detailed systematic study. For each series we start from our ``golden case'' and incrementally vary the properties of the binary -- the total mass, mass ratio, \gls{SNR}, inclination angle -- one at a time.

To prevent prior railing
we only consider simulated \gls{IMBH} binaries with total masses $200 \leq M \leq 450\, M_{\odot}$ and mass ratios $0.25 \leq q \leq 1$. Owing to the lack of observation evidence and knowledge for the formation mechanisms of light \gls{IMBH} binaries, we did not assume an astrophysical mass and spin distribution for our injections. We instead considered a uniform distribution.

When presenting the results for each analysis, we focus on the inferred total mass ($M$), mass ratio ($q$), spin magnitudes ($\chi_1$ and $\chi_2$), and inclination angle ($\theta_{\mathrm{JN}}$),
as they have the broadest astrophysical relevance.

\subsubsection{Scaling Total Mass Analyses} \label{sec:total_mass_series}

In general, as the total mass of the binary increases templates starting at larger frequencies diverge strongly from the injected value, see Fig.~\ref{fig:total_mass_posteriors}. Interestingly, we see that for $M \lesssim 250 \, M_{\odot}$, biases from model starting frequency are negligible: across all parameters, the posterior distributions for different template starting frequencies agree well and peak near the injected values. We suspect this is because all higher order multipole content is incorporated in the template before the \gls{ISCO} frequency of the binary, see Eq.~\ref{eq:Kerr_ISCO_freq} in Appendix~\ref{sec:appendix_notation}. 
For $M = 350\, M_{\odot}$, the \gls{ISCO} frequency is $\sim 35\, \mathrm{Hz}$. This means that for templates with $f_{22} = 20 \, \mathrm{Hz}$, we are missing $\gtrsim 5\, \mathrm{Hz}$ of \gls{GW} content in the merger and ringdown.
In contrast, the \gls{ISCO} frequency for $M = 450\, M_{\odot}$ reduces to $\sim 28\, \mathrm{Hz}$ implying that for templates with $f_{22} = 13\, \mathrm{Hz}$, the $(\ell, m) = (4, 4)$ multipole is only just turning on and artefacts from the Fourier transform may influence results.

Focusing specifically on the inferred total mass, when the template starts at $f_{22} = 20 \, \mathrm{Hz}$ our analyses underestimate the true total mass of the source for $M = 300\, M_{\odot}$ by $\sim 15\, M_{\odot}$ and overestimate the injected value for $M \gtrsim 350 , M_{\odot}$ by $\gtrsim 20\, M_{\odot}$. For binaries with $M > 400\, M_{\odot}$, the true total mass of the binary is recovered within the $\sim 0.4\%$ credible interval. This suggests that for binaries with total mass $M > 250\, M_{\odot}$, missing higher order multipole content between $20 - 40\, \mathrm{Hz}$ leads to biased estimates in the total mass of the source. When the template starts at $f_{22} = 13\, \mathrm{Hz}$, we generally recover comparable posterior distributions to the $f_{22} = 10\, \mathrm{Hz}$ cases.
This suggests that missing power from the $(\ell, m) = (3, 3)$ multipole is the cause of the biased measurements.
Only for masses $M \gtrsim 450 \, M_{\odot}$ does the $f_{22} = 13\, \mathrm{Hz}$ diverge strongly from the $f_{22} = 10\, \mathrm{Hz}$ analysis, indicating that from this mass the $(\ell, m) = (4, 4)$ multipole becomes important.

The inferred primary spin magnitude shows very similar trends to the total mass: we see overestimated spins for binaries with total mass $M \gtrsim 350 \, M_{\odot}$ and templates with starting frequency $f_{22} = 20 \, \mathrm{Hz}$. We also see an underestimated primary spin magnitude for templates with $f_{22} = 13 \, \mathrm{Hz}$ for binaries with $M \gtrsim 450 \, M_{\odot}$. This is expected since at this mass ratio, the primary spin magnitude dominates the morphology of the \gls{GW} signal, and both the total mass and spin characterise the merger and ringdown of the template.
Interestingly, for templates with $f_{22} = 20\, \mathrm{Hz}$ the secondary spin incorrectly rails against maximum black hole spin for high mass binaries. 

When focusing on the inferred mass ratio, we see that template starting frequency has little affect on the inferred distribution for the majority of cases. The exceptions are binaries with $M = 300\, M_{\odot}$ and $M = 450\, M_{\odot}$\footnote{Note that for templates with $f_{22} = 13$ and $20\, \mathrm{Hz}$, the $M = 450\, M_{\odot}$ case rails against the lower bound of the mass ratio prior.} where we observe a systematic bias toward more asymmetric component masses.

The inclination angle recoveries remain consistently well constrained across the mass range. A deviation is only seen in the $f_{22} = 20 \, \mathrm{Hz}$ case, which begins to underestimate for $350 \, M_{\odot} \lesssim M \lesssim 450 \, M_{\odot}$.
This behaviour is expected because at high \gls{SNR} ($\rho = 75$), the well-known $d_{\mathrm{L}}-\theta_{\mathrm{JN}}$ degeneracy is broken~\citep{Usman:2018imj,Kalaghatgi:2019log}, leading to robust constraints on the inclination angle

Across all five parameters, the $M = 300 \, M_{\odot}$ case stands out as unique. 
For this simulation, the $f_{22} = 13 \, \mathrm{Hz}$ result is noticeably less accurate than at adjacent masses, more closely resembling the behaviour of $M = 450 \, M_{\odot}$.
This indicates that omitting the $(4,4)$ multipole at this specific configuration has a stronger impact on recovery than expected. To rule out the possibility of sampling issues, we reran the total-mass series with an increased number of live points (2000 and 3000 compared to 1000 used by default) and obtained consistent results.

\begin{figure}
    \includegraphics[width=0.5\textwidth]{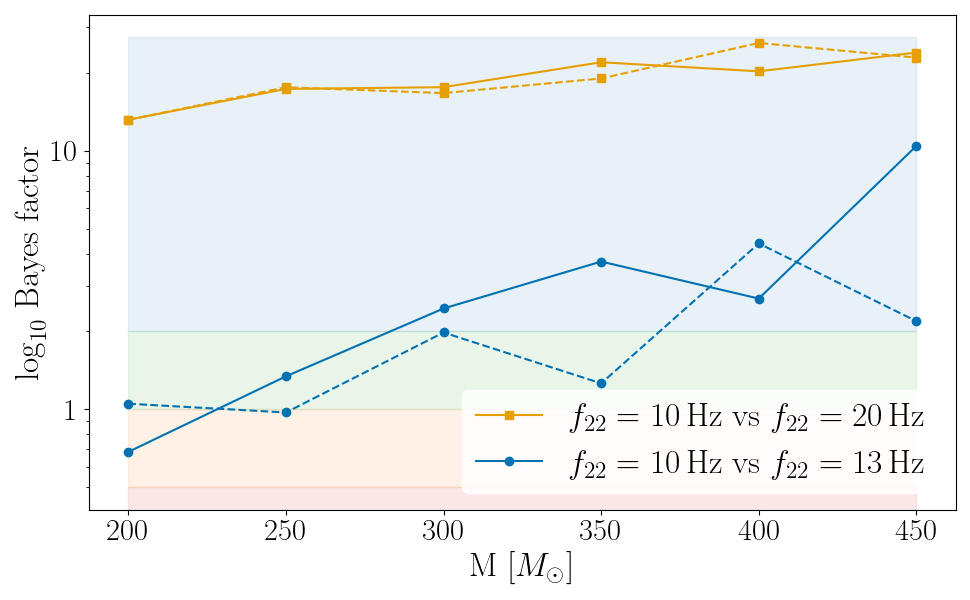}
    \caption{log$_{10}$ Bayes factors comparing analyses with starting frequencies $f_{22} = 13 \, \mathrm{Hz}$ and $f_{22} = 20 \, \mathrm{Hz}$ against $f_{22} = 10 \, \mathrm{Hz}$ across the total mass series. Higher Bayes factors indicate stronger preference for the lower starting frequency analysis. In solid we show the Bayes factors for binaries with spin magnitudes $\chi_{i} = 0.7$ and in dashed we show the Bayes factors for binaries with spin magnitudes $\chi_{i} = 0.9$. Given that the $f_{22} = 10 \, \mathrm{Hz}$ generally well recovers the injected parameters, larger Bayes factors highlight increasing bias in parameter recovery as the total mass increases. The red, orange, green, and blue shaded regions indicate no substantial, substantial, strong, and decisive evidence in favour of the $f_{22} = 10 \, \mathrm{Hz}$ analysis~\citep{Kass:1995loi}.}
    \label{fig:total_mass_bf}
\end{figure}

As shown in Fig.~\ref{fig:total_mass_bf}, the Bayes factor for $f_{22} = 10 \, \mathrm{Hz}$ versus $f_{22} = 13 \, \mathrm{Hz}$ crosses the $\log_{10} \mathcal{B} = 2$ threshold between $250 \, M_{\odot}$ and $300 \, M_{\odot}$. This indicates that for \gls{IMBH} systems with total masses $> 250 \, M_{\odot}$, inclusion of the $(4,4)$ multipole is necessary to avoid significant biases in parameter estimation. This threshold is considerably lower than the previously indicated value of $450 \, M_{\odot}$, because the Bayes factor accounts for the full 15-dimensional parameter space. The Bayes factor shows a small decrease at $M = 400 \, M_{\odot}$, reflecting the improved accuracy at this mass, when compared to the $M = 300 ,\ M_{\odot}$, seen in Fig.~\ref{fig:total_mass_posteriors}.

\begin{figure*}
    \includegraphics[width=0.92\textwidth]{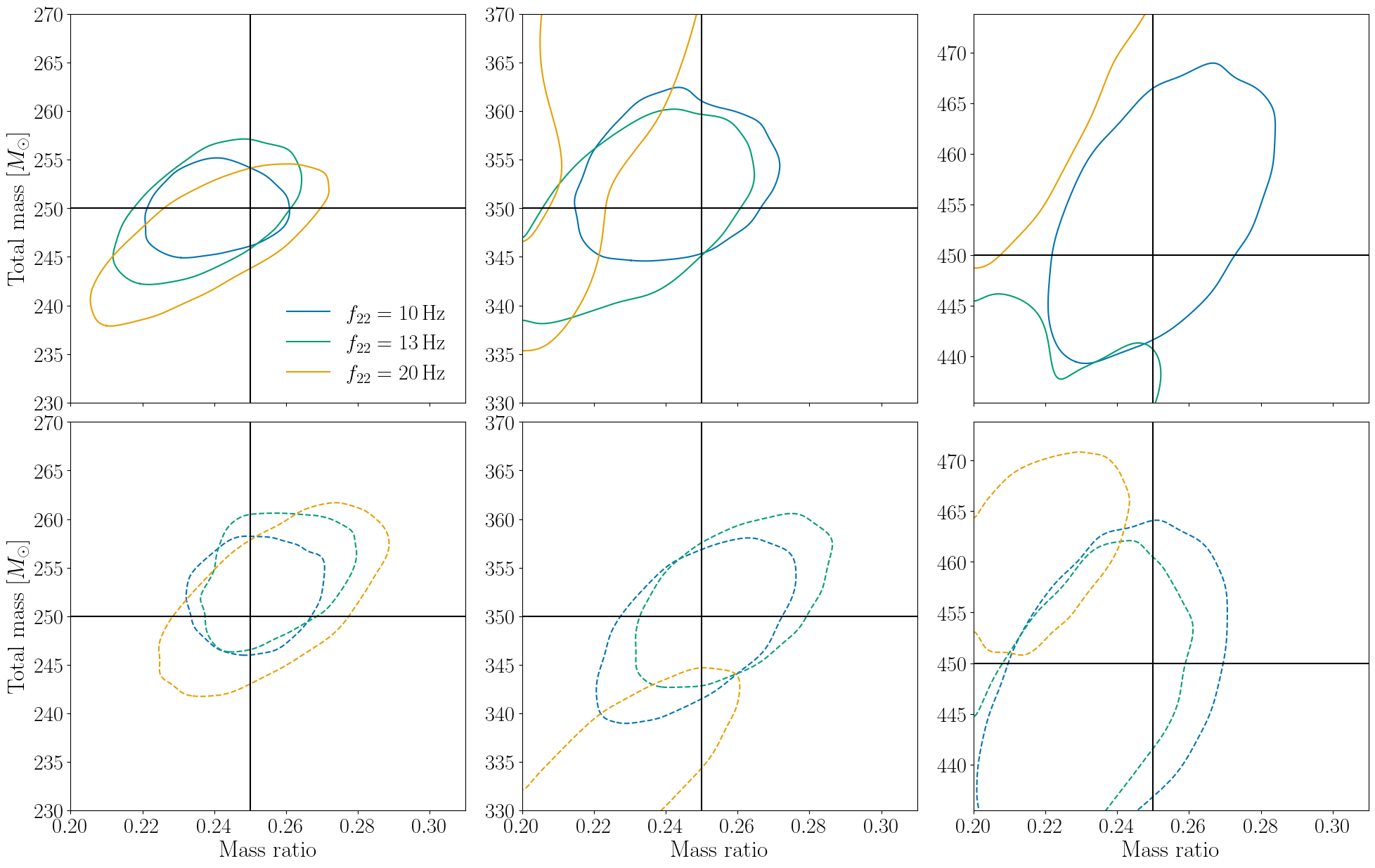}
    \caption{Two-dimensional marginalised posterior distribution for the inferred total mass $M$ and mass ratio $q$ when varying the total mass of injection. In blue, orange and green we show the posterior distribution obtained when the template starting frequency is $f_{22} = 10, 13$ and $20\, \mathrm{Hz}$ respectively. In the top row we show results for binaries with spin magnitude $\chi_{i} = 0.7$, and in the bottom row we show results for binaries with $\chi_{i} = 0.9$. The left, middle and right columns show the results for injection with injected total mass $M = 250 \, M_{\odot}, 350\, M_{\odot}$, and $450\, M_{\odot}$ respectively. In all panels, the black solid vertical and horizontal lines show the true source properties of the binary and contours represent the 90\% credible interval.}
    \label{fig:total_mass_high_spin_posteriors}
\end{figure*}

\subsubsection{The impact of spin magnitude and spin-precession} \label{subsubsec:high_spin_tilt}

GW190521~\citep{LIGOScientific:2020iuh} and GW231123~\citep{LIGOScientific:2025rsn} provide the most compelling evidence for \gls{IMBH} binaries below $10^{3}\, M_{\odot}$. Bayesian analyses of GW231123 in particular show that both black holes are likely rapidly spinning, and the binary may have been precessing~\citep{LIGOScientific:2025rsn}. As a result, we next examine how our conclusions change for black holes with higher spin magnitudes and large in-plane spins. We repeat the same injections as in Sec.~\ref{sec:total_mass_series} but now increase the spin magnitudes to $\chi_1 = \chi_2 = 0.9$, and increase the spin tilts to $\theta_{1} = 0.8, \,\theta_{2} = 1.5\, \mathrm{rad}$.

When increasing the spin magnitudes, we observe the same overall trends as before, see Fig.~\ref{fig:total_mass_high_spin_posteriors}: as the total mass of the binary increases, the starting frequency of the template becomes increasingly important in the recovery. We see significant biases in the inferred total mass and mass ratio for injections with $M \gtrsim 350 \, M_{\odot}$ for templates with starting frequency $f_{22} = 20 \, \mathrm{Hz}$. Interestingly, we see that the difference between templates with starting frequency $f_{22} = 10 \, \mathrm{Hz}$ and $f_{22} = 13 \, \mathrm{Hz}$ is noticeably smaller for higher spin configurations.

The Bayes factors reflect this behaviour, yielding lower values for the $f_{22} = 13 \, \mathrm{Hz}$ versus $f_{22} = 10 \, \mathrm{Hz}$ comparison than in the original total-mass series. In fact, for this higher-spin case the Bayes factor only crosses the threshold $\log_{10} \mathcal{B} > 2$ at $M \gtrsim 350 \, M_{\odot}$, see Fig.~\ref{fig:total_mass_bf}. This indicates that higher spins reduce the impact of missing multipoles at moderate masses, allowing the $(4,4)$ multipole to be neglected in a wider range of systems.

When increasing the degree of spin-precession in the binary, we observe a larger bias in the inferred parameters. The Mahalanobis recovery scores for $f_{22} = 20 \, \mathrm{Hz}$ were zero for $M \gtrsim 300 \, M_{\odot}$ (see Appendix~\ref{sec:appendix_mahal_recov_tables} for the full table of scores) indicating that the injected value remained outside the 90\% credible interval for all projections.
The Bayes factor results differ notably from the regular total-mass series. When comparing the $f_{22} = 10 \, \mathrm{Hz}$ and $f_{22} = 13 \, \mathrm{Hz}$ analyses, $M = 300 \, M_{\odot}$ is the only case where $\log_{10} \mathcal{B} < 2$. This suggests that the $(4,4)$ multipole is essential for all masses except $M = 300 \, M_{\odot}$ when the binary exhibits significant spin-precession.

To see if there was a true correlation between the degree of spin-precession in the binary and the level of bias in the inferred parameters we injected \glspl{GW} from binaries with spins aligned with the orbital angular momentum. For these injections we obtained the same generic behaviour as before: the bias from model starting frequency generally increases with total mass.
This suggests that the relative importance of the $(4,4)$ multipole remains small across most of the mass range, and that lowering the spin-tilt angles does not significantly affect the overall accuracy of parameter recovery.

\subsubsection{Scaling Mass Ratio Analyses}
\begin{figure*}
    \includegraphics[width=0.92\textwidth]{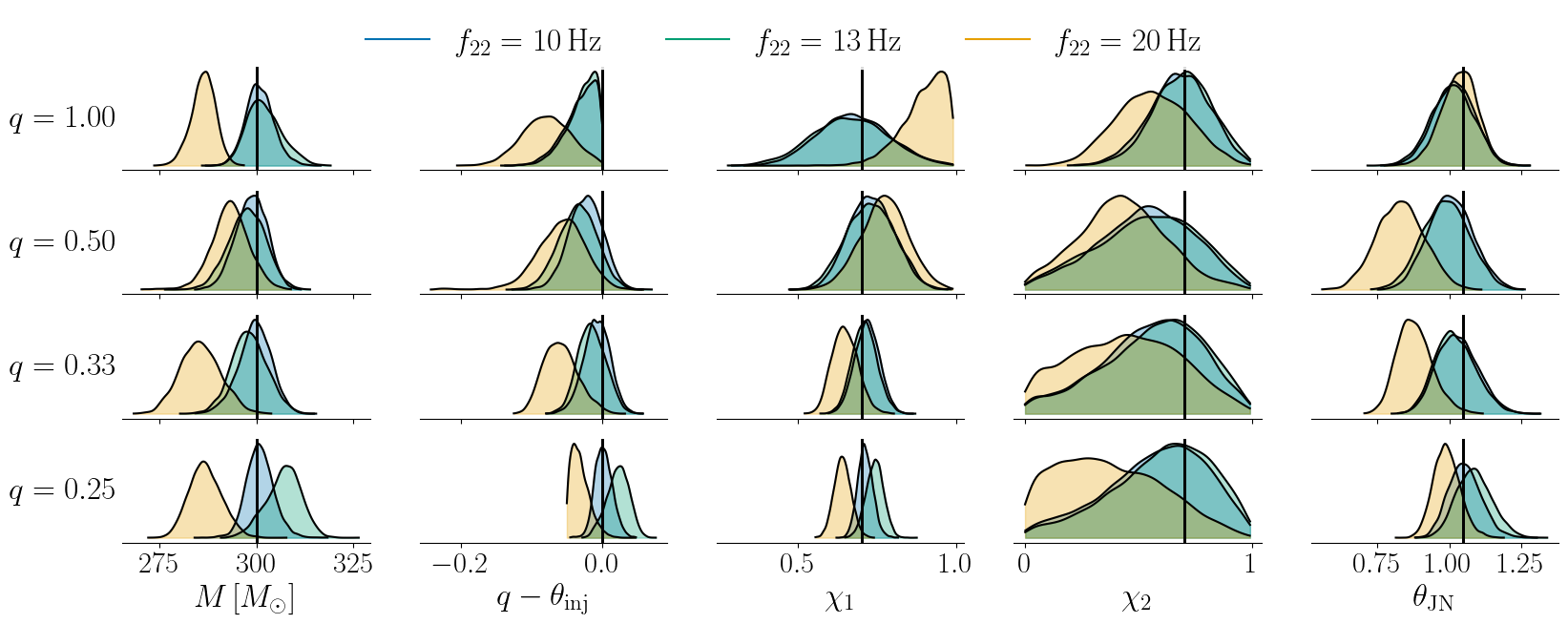}
    \caption{Same as Fig.~\ref{fig:total_mass_posteriors} except we show the results when varying the mass ratio of the binary. For the second column we shift the posterior distribution by the injected mass ratio to centre the true value around $0$. The different rows shows the results for different mass ratio injections. For the $q = [1, 0.25]$ cases, prior railing prevents the posterior from returning a Gaussian-like distribution.}\label{fig:mass_ratio_posteriors}
\end{figure*}

For equal-mass systems ($Q = m_{1} / m_{2} = 1$), recovery of all parameters except the inclination angle is poor with $f_{22} = 20 \, \mathrm{Hz}$, see Fig.~\ref{fig:mass_ratio_posteriors}. Equal-mass binaries have intrinsically shorter inspirals than asymmetric systems of the same total mass~\citep{Cutler:1994ys}, and hence starting the template at higher frequencies leaves insufficient \gls{GW} content to constrain parameters. Similarly, the absence of information from the $(3,3)$ or $(4,4)$ multipoles further prevents ruling out asymmetric mass ratios.
At $Q = 2$, biases are reduced across most parameters, with all three templates performing comparably. Here, the inspiral is long enough to provide sufficient information in the band, while the asymmetry is still too weak for higher multipoles to dominate.

The recovery of $M$, $q$, $\chi_1$, and $\chi_2$ follows a common trend. Templates with $f_{22} = 20 \, \mathrm{Hz}$ generally underestimate the injected values, with the strongest biases at $Q = 1$, $Q = 3$, and $Q = 4$. In contrast, the $f_{22} = 13 \, \mathrm{Hz}$ performs comparably to $f_{22} = 10 \, \mathrm{Hz}$, with templates recovering the true values across most configurations, with deviations only at $Q = 4$. Between the spin parameters, $\chi_1$ is more reliably constrained than $\chi_2$, consistent with the greater influence of the primary black hole on the waveform.

For templates with starting frequency $f_{22} = 20 \, \mathrm{Hz}$, the inferred inclination angle remains accurate for \gls{IMBH} sources with $Q = 1$. This deteriorates at $Q = 2$, and then gradually improves for $Q = 3$--$4$. Notably, this is the only parameter for which all three starting frequencies do not perform similarly at $Q = 2$. The high accuracy at $Q = 1$ and the trend seen as $Q$ increases suggests that, for symmetric and highly asymmetric systems, the recovery of the inclination angle is more robust and generally more accurate compared to moderately asymmetric configurations.

\begin{figure*}
    \includegraphics[width=0.92\textwidth]{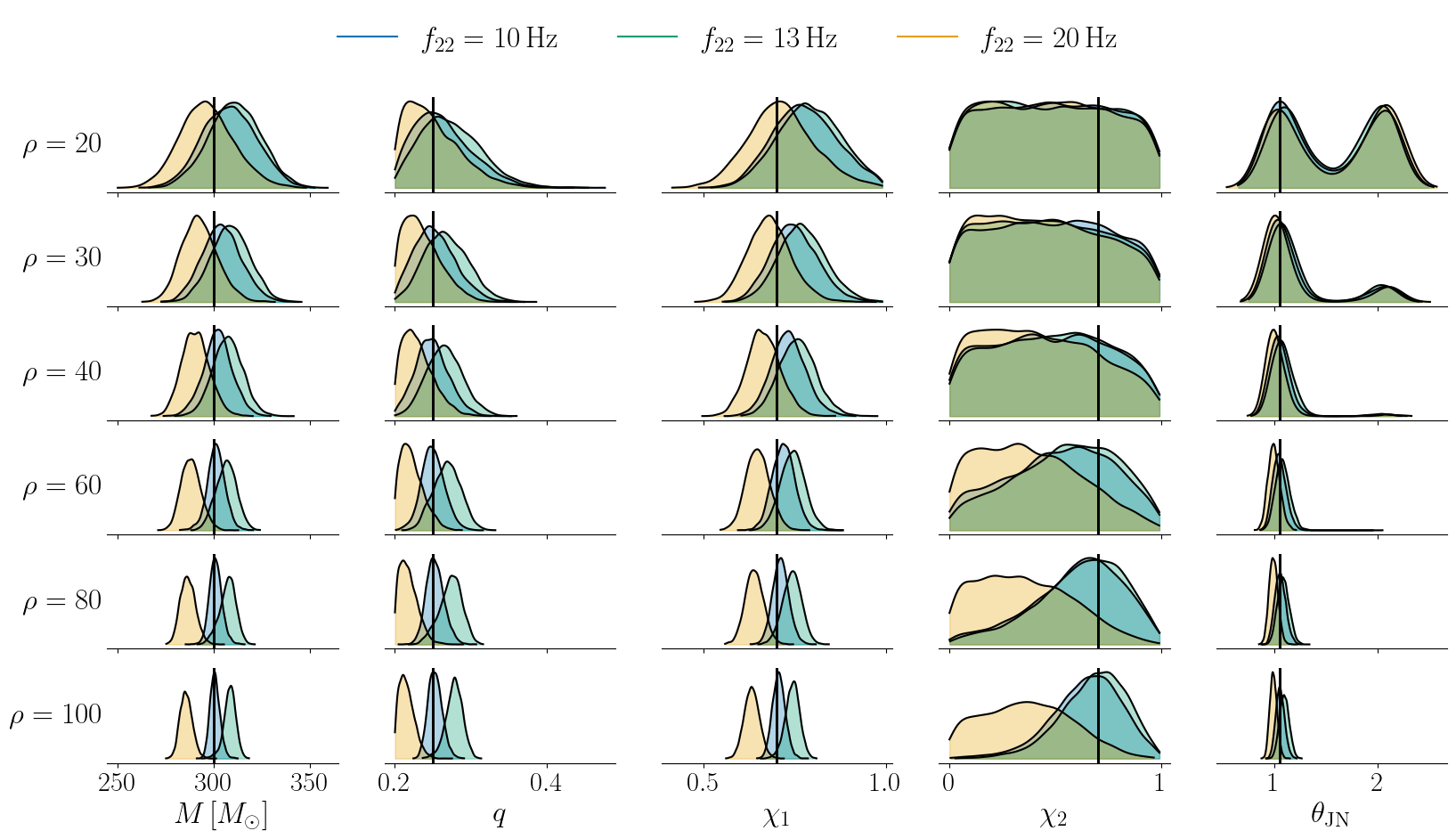}
    \caption{Same as Fig.~\ref{fig:total_mass_posteriors} except we show the results when varying the \gls{SNR} of the observed signal. Owing to the size of this series and the clarity of the observed trends, we omit some intermediate \gls{SNR} steps for simplicity.}
    \label{fig:snr_posteriors}
\end{figure*}

The trend changes notably at $Q = 4$, where templates with starting frequency $f_{22} = 13\, \mathrm{Hz}$ begin to deviate from the injected values across multiple parameters. This transition indicates that the $(4, 4)$ multipole becomes increasingly important for accurate parameter recovery in highly asymmetric systems, consistent with~\cite{Mills:2020thr}.

The Bayes factors follow the trends in Fig.~\ref{fig:mass_ratio_posteriors}. Differences between $f_{22} = 10 \, \mathrm{Hz}$ and $f_{22} = 13 \, \mathrm{Hz}$ remain modest ($\log_{10} \mathcal{B} < 5$) for all mass ratios, consistent with the $(4, 4)$ multipole being subdominant but still relevant. At $Q = 3$, the Bayes factor drops below $\log_{10} \mathcal{B} = 2$, suggesting that moderately asymmetric systems are least sensitive to missing $(4, 4)$ content. In contrast, comparing $f_{22} = 10 \, \mathrm{Hz}$ to $f_{22} = 20 \, \mathrm{Hz}$ shows a non-monotonic trend: the Bayes factor peaks at $Q = 1$, decreases to a minimum near $Q = 3$, and rises again toward $Q = 4$. This suggests that biases from higher starting frequencies are most severe for equal-mass and highly asymmetric binaries, while $Q \sim 2$ lies in a bias-resistant regime.

\subsubsection{Scaling \gls{SNR} Analyses} \label{sec:snr_series}

We observe a clear monotonic relationship between increasing $\rho$ and the severity of bias from template starting frequency. As shown in Fig.~\ref{fig:snr_posteriors}, all three starting frequencies perform comparably at low \gls{SNR} ($\rho \simeq 20$), with posterior distributions largely overlapping and recovering the injected values. At higher \glspl{SNR}, the $f_{22} = 13 \, \mathrm{Hz}$ and $f_{22} = 20 \, \mathrm{Hz}$ cases progressively narrow, revealing a divergence from the $f_{22} = 10 \, \mathrm{Hz}$ baseline, making systematic biases increasingly apparent as statistical uncertainties reduce.

\begin{figure}
    \includegraphics[width=0.45\textwidth]{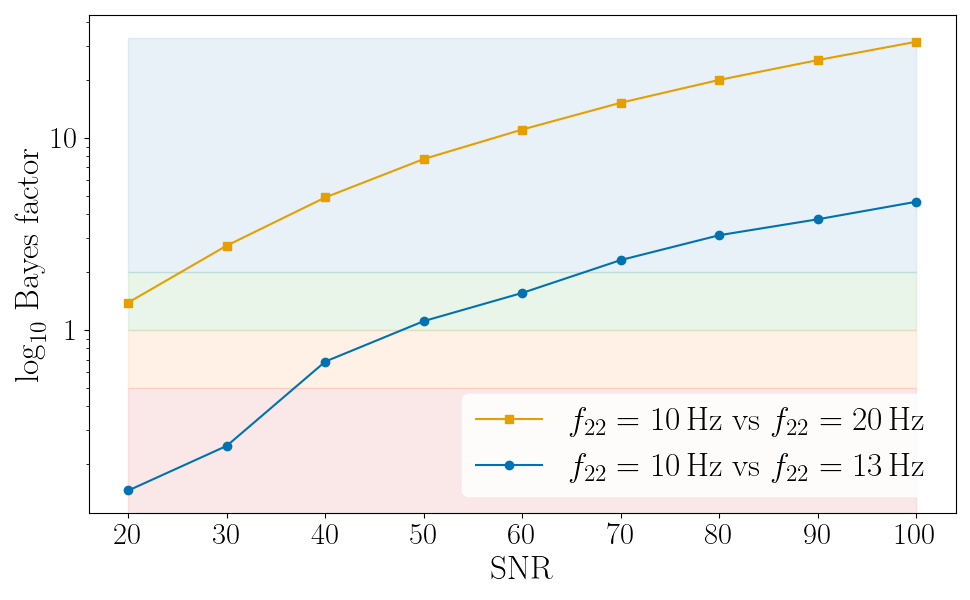}
    \caption{Same as Fig.~\ref{fig:total_mass_bf} except we show the results when varying the signal-to-noise ratio of the binary.}
    \label{fig:snr_bf}
\end{figure}

\begin{figure*}
    \includegraphics[width=0.92\textwidth]{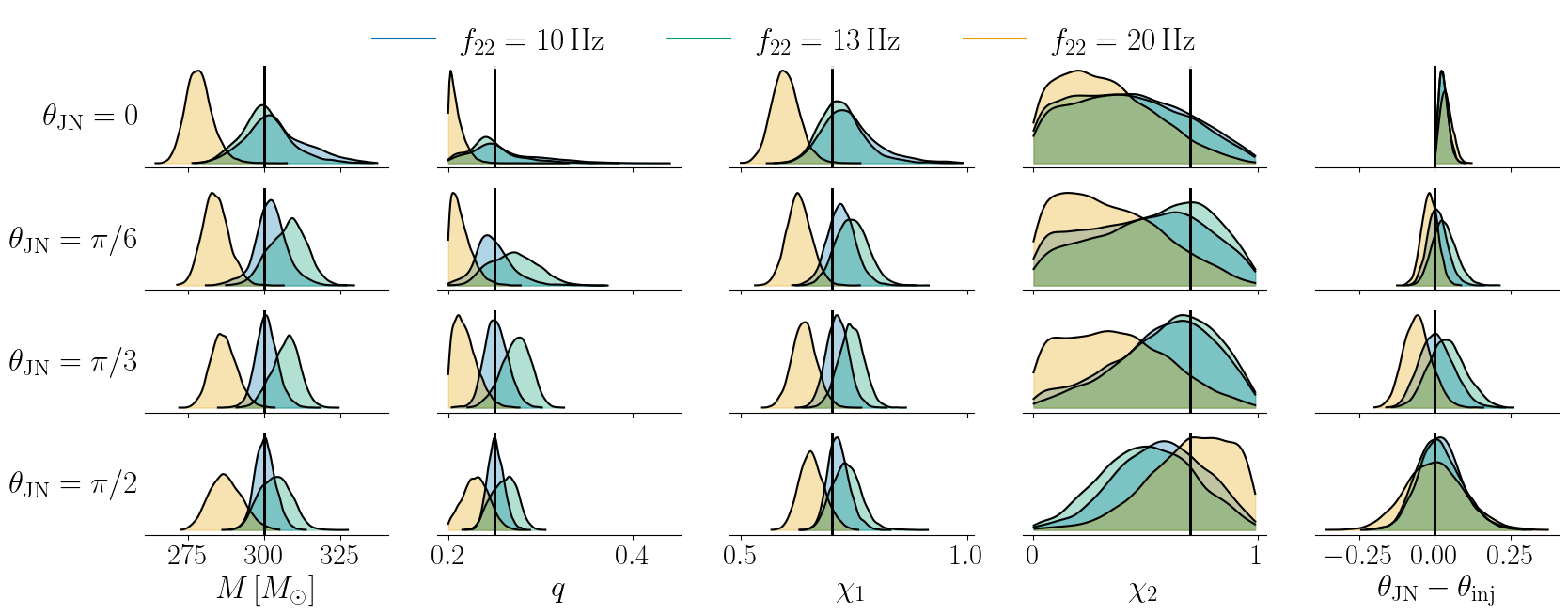}
    \caption{Same as Fig.~\ref{fig:total_mass_posteriors} except we show now the results when varying the inclination angle of the binary.}
    \label{fig:inclination_angle_posteriors}
\end{figure*}

For most marginalised distributions, we observe a posterior that underestimates the injected value for templates with starting frequency $f_{22} = 20 \, \mathrm{Hz}$ at \glspl{SNR} greater than $\rho = 30$. By contrast, templates with $f_{22} = 13 \, \mathrm{Hz}$ display the opposite behaviour: we observe systematic biases overestimating the injected values, with the exception of the secondary spin magnitude $\chi_2$, which closely matches the $f_{22} = 10 \, \mathrm{Hz}$ baseline across the full series. This suggests that the $(4,4)$ multipole has little to no impact on recovering $\chi_2$, regardless of the \gls{SNR}. However, the $(\ell, m) = (4, 4)$ multipole is needed to accurately recover the $M$, $q$, and $\chi_{1}$ for $\rho \gtrsim 70$.

The inclination angle recovery shows very little deviation from the injected value for any of the three starting frequencies. At low \gls{SNR} ($\rho \lesssim 30$), the posteriors are broad and poorly constrained, exhibiting the expected bimodal structure.
As the \gls{SNR} increases beyond $\rho \gtrsim 40$, the posterior distributions gradually tighten, though extended posterior tails remain until $\rho \gtrsim 70$. Importantly, across the full \gls{SNR} range all three starting frequencies perform almost identically, reinforcing the conclusion that the inclination angle is determined mainly by the signal’s overall amplitude and polarisation structure, rather than the starting frequency of higher-order multipoles.

This conclusion is supported by the Bayes factor results, which show a clear monotonic increase for both the $f_{22} = 10 \, \mathrm{Hz}$ vs $f_{22} = 13 \, \mathrm{Hz}$ and $f_{22} = 10 \, \mathrm{Hz}$ vs $f_{22} = 20 \, \mathrm{Hz}$ comparisons, as illustrated in Fig.~\ref{fig:snr_bf}. In particular, the $f_{22} = 10 \, \mathrm{Hz}$ vs $f_{22} = 20 \, \mathrm{Hz}$ case demonstrates that for $\rho > 20$, fully including the $(3,3)$ multipole is essential for accurate recovery. Similarly, the $f_{22} = 10 \, \mathrm{Hz}$ vs $f_{22} = 13 \, \mathrm{Hz}$ comparison shows that fully including the $(4,4)$ multipole becomes necessary only at very high detection significance, $\rho \gtrsim 70$. In Table~\ref{tab:missing_snr} we compare the recovered and missing orthogonal \glspl{SNR} for this series. We see that our \gls{SNR} thresholds for choosing a starting frequency that fully includes the $(3, 3)$ and $(4, 4)$ multipoles map to an average missing orthogonal \gls{SNR} $\gtrsim 5$.

\subsubsection{Scaling Inclination Angle Analyses} \label{subsubsec:inclination_angle}

As the inclination-angle of the binary is varied, templates with $f_{22} = 20 \, \mathrm{Hz}$ consistently underestimate the injected values across $M$, $Q$, $\chi_{1}$, and $\chi_{2}$, as seen in Fig.~\ref{fig:inclination_angle_posteriors}. A notable exception occurs for the secondary spin $\chi_{2}$ at $\theta_{\mathrm{JN}} = \pi / 2\, \mathrm{rad}$. Here, the $f_{22} = 20 \, \mathrm{Hz}$ case instead overestimates the injected value. More generally, all starting frequencies struggle to constrain $\chi_{2}$ in edge-on and face-on systems, reflecting the fundamental difficulty of measuring subdominant spin contributions.

Templates with starting frequency $f_{22} = 13 \, \mathrm{Hz}$ demonstrate excellent agreement with the $f_{22} = 10 \, \mathrm{Hz}$ analysis for both face-on and edge-on configurations across total mass, mass ratio, and primary spin parameters. However, their performance degrades at intermediate inclinations ($\theta_{\mathrm{JN}} = \pi / 6$ and $\pi / 3\, \mathrm{rad}$), where they consistently overestimate the injected values. This behaviour suggests that the $(4, 4)$ multipole becomes increasingly important for accurate parameter recovery at intermediate orientations, where the higher order multipoles start to become more important. Interestingly, the improvement in the $f_{22} = 13 \, \mathrm{Hz}$ case at $\theta_{\mathrm{JN}} = \pi / 2\, \mathrm{rad}$ is unexpected as the $(2, 2)$ should be minimised here, while the $(3, 3)$ and $(4, 4)$ are maximised~\citep{Mills:2020thr}.

\begin{table}
    \centering
    \begin{tabular}{|c|c|c|c|}
        \toprule
        Injected & $f_{22}\, [\mathrm{Hz}]$ & Recovered & Missing \\
        \toprule
        \multirow{2}{*}{20} & 13 & $19.7^{+0.2}_{-0.2}$ & 3.6 \\
        & 20 & $19.5^{+0.2}_{-0.2}$ & 4.3 \\
        \midrule
        \multirow{2}{*}{30} & 13 & $29.8^{+0.1}_{-0.2}$ & 3.8 \\ 
        & 20 & $29.6^{+0.1}_{-0.2}$ & 5.0 \\ 
        \midrule
        \multirow{2}{*}{40} & 13 & $39.8^{+0.1}_{-0.1}$ & 4.0 \\
        & 20 & $39.6^{+0.1}_{-0.1}$ & 5.9 \\
        \midrule
        \multirow{2}{*}{60} & 13 & $59.8^{+0.1}_{-0.1}$ & 4.6 \\
        & 20 & $59.46^{+0.05}_{-0.08}$ & 8.0 \\
        \midrule
        \multirow{2}{*}{80} & 13 & $79.83^{+0.05}_{-0.07}$ & 5.3 \\
        & 20 & $79.33^{+0.04}_{-0.06}$ & 10.4 \\
        \midrule
        \multirow{2}{*}{100} & 13 & $99.83^{+0.04}_{-0.06}$ & 5.8 \\
        & 20 & $99.19^{+0.03}_{-0.05}$ & 12.7 \\
        \bottomrule
    \end{tabular}
    \caption{Comparison between the injected, recovered and average missing (orthogonal) \gls{SNR} for the analyses described in Sec.~\ref{sec:snr_series}. For simplicity, we omit some intermediate \gls{SNR} steps, as in Fig.~\ref{fig:snr_posteriors}.}
    \label{tab:missing_snr}
\end{table}

The inclination angle exhibits the most robust recovery performance across all starting frequencies and orientations, with edge-on and face-on systems performing best. The three starting frequencies yield nearly identical recovery for $\theta_{\mathrm{JN}}$ across all orientations, with only a minor deviation at $\theta_{\mathrm{JN}} = \pi / 3\, \mathrm{rad}$ in the $f_{22} = 20 , \mathrm{Hz}$ case. For this case we observe a slight underestimation, and the posteriors widen as the angle increases. This behaviour is mostly consistent with expectations, since face-on configurations are dominated by the $(2,2)$ multipole and are therefore easiest to constrain, while for edge-on systems the higher order multipoles play a more important role~\citep{Mills:2020thr}.

The Mahalanobis recovery scores provide quantitative confirmation of the trends observed in the one dimensional marginalised posterior distributions (see Appendix~\ref{sec:appendix_mahal_recov_tables} for the full table of scores). Templates with starting frequency $f_{22} = 20 \, \mathrm{Hz}$ demonstrate significantly degraded performance at face-on ($0\, \mathrm{rad}$), moderately inclined ($\pi / 6\, \mathrm{rad}$), and highly inclined ($\pi / 3\, \mathrm{rad}$) orientations. This systematic underperformance across inclination angles suggests the $(3, 3)$ multipole is important for accurate parameter estimation in \gls{IMBH} systems, regardless of the binary's orientation. When calculating Bayes factors in favour of the $f_{22} = 10\, \mathrm{Hz}$ over the $f_{22} = 13\, \mathrm{Hz}$ analysis, we see that the $\log_{10} \, \mathcal{B} = 2$ threshold is crossed between $\theta_{\mathrm{JN}} = \pi / 6\, \mathrm{rad}$ and $\pi / 3\, \mathrm{rad}$. This confirms that the $(4, 4)$ multipole becomes important for higher inclination angles.

\section{GW231123\_135430} \label{sec:GW231123}

GW231123 was observed at \gls{SNR} $\rho=22.6^{+0.2}_{-0.3}$. The properties of the source varied depending on the model used for inference, but for \textsc{NRSur7dq4} the total mass (in the detector-frame), mass ratio and spin magnitudes were inferred to be: $M = 320^{+10}_{-30}\, M_{\odot}$, $q = 0.86^{+0.14}_{-0.11}$\ $\chi_{1} = 0.89^{+0.11}_{-0.20}$ and $\chi_{2} = 0.91^{+0.09}_{-0.19}$ respectively~\citep{LIGOScientific:2025rsn}. Since GW231123 is consistent with the injections performed in this study, we additionally assess the impact of model starting frequency on GW231123's inferred source properties. We used the same priors, sampler settings, power spectral densities and
calibration envelopes as those described in~\cite{LIGOScientific:2025rsn}. To ensure consistency between analyses, we used a reference frequency of $20\, \mathrm{Hz}$ rather than $10\, \mathrm{Hz}$ as used in~\cite{LIGOScientific:2025rsn}. 

\begin{figure}
    \includegraphics[width=0.45\textwidth]{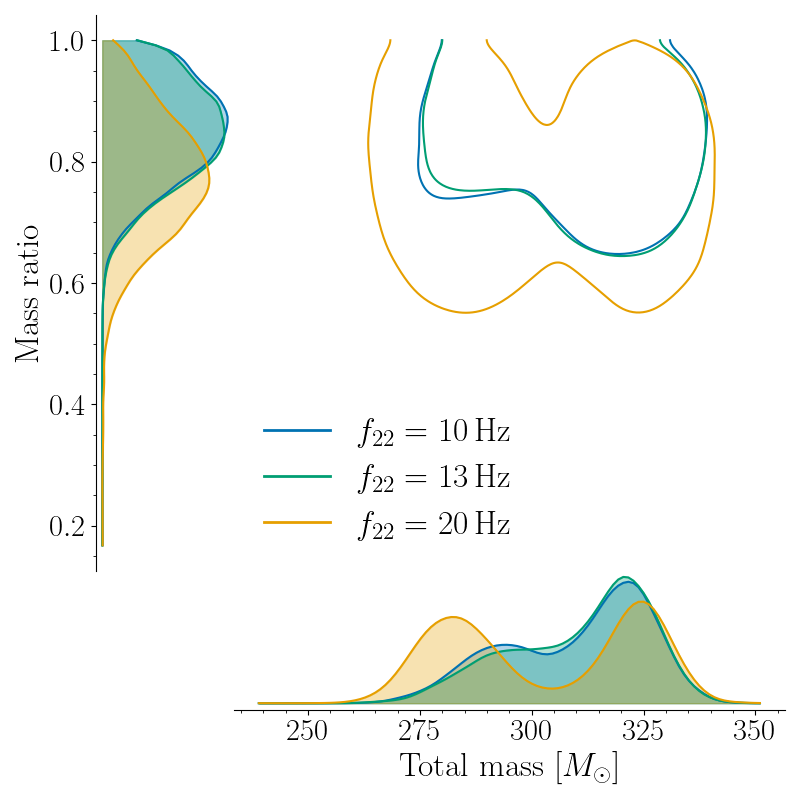}
    \caption{The two-dimensional marginalized posterior distribution for the inferred total mass $M$ and mass ratio $q$ for our re-analysis of GW231123\_135430. In blue, green and orange we show the posterior distribution obtained when the template starting frequency is $f_{22} = 10$, $13$, and $20 \, \mathrm{Hz}$ respectively. The contours represent the inferred 90\% credible interval}
    \label{fig:gw231123}
\end{figure}

In Fig.~\ref{fig:gw231123}, we see that GW231123's source properties vary depending on the starting frequency of the template. Although the inferred 90\% credible interval remains consistent, when the starting frequency is $f_{22} = 20\, \mathrm{Hz}$ the missing power from higher order multipoles causes the total mass to become bimodal with more support for asymmetric masses. As expected, we obtain comparable results between the $f_{22} = 10\, \mathrm{Hz}$ and $f_{22} = 13\, \mathrm{Hz}$ analyses. This indicates that the $(\ell, m) = (3, 3)$ multipole contributes to the total signal power between $20 - 30\, \mathrm{Hz}$, while there is little evidence for the $(4, 4)$. Although not shown here, we see that the inferred spin magnitudes remain agnostic to the template starting frequency. In terms of Bayes factors, we see that the $f_{22} = 10\, \mathrm{Hz}$ analysis is preferred over the $f_{22} = 13\, \mathrm{Hz}$ and $f_{22} = 20\, \mathrm{Hz}$ by $\log_{10} \, \mathcal{B} = 0.3$ and $2.2$ respectively.

In Sec.~\ref{sec:snr_series}, we highlighted that \gls{GW} signals observed with $\rho < 20$, a starting frequency of $20\, \mathrm{Hz}$ can safely be used as statistical uncertainties dominate. We see that our results in real \gls{GW} noise remain consistent with this conclusion.

\section{Discussion} \label{sec:discussion}

In this work, we present a detailed systematic study investigating how the starting frequency of time-domain models impacts Bayesian parameter estimation for light \gls{IMBH} binaries. Using a comprehensive set of injection and recovery analyses, we quantify biases associated with the ``missing multipole problem" for binaries with total masses: $M = 200$ to $450 \, M_{\odot}$, mass ratios: $q = 0.25$ to $1$, \glspl{SNR}: $\rho = 20$ to $100$, and inclination angles: $\theta_{\mathrm{JN}} = 0$ to $\pi / 2 \, \mathrm{rad}$.
Our results show that the choice of template starting frequency can produce significant biases in parameter estimation, particularly for total masses $\gtrsim 300 \, M_{\odot}$ and detections at \glspl{SNR} $\gtrsim 70$. The magnitude of the bias depends strongly on the astrophysical properties of the system, with the most massive, highly asymmetric, and edge-on binaries being the most susceptible to systematic errors.

Based on our results we suggest that when the likelihood is integrated from $f_{\mathrm{low}} = 20 \, \mathrm{Hz}$, templates can be started from $f_{22} = 20 \, \mathrm{Hz}$ for \glspl{SNR} $\lesssim 20$ as statistical uncertainties naturally encompass potential biases due to model starting frequency. For \gls{SNR} values $20 \lesssim \rho \lesssim 70$, the $(3, 3)$ multipole should be included by using $f_{22} = 13 \, \mathrm{Hz}$ or lower. For \gls{SNR} $\gtrsim 70$, the $(4, 4)$ multipole must also be incorporated using $f_{22} = 10 \, \mathrm{Hz}$. At $\mathrm{SNR} \sim 75$, the $(4, 4)$ multipole becomes essential for systems with total masses above $250 \, M_{\odot}$, for all mass ratios except the moderately asymmetric case of $q \approx 0.33$, and for inclination angles of $\pi/3$ and $\pi/2 \, \mathrm{rad}$. At inclinations ($0$ and $\pi/6 \, \mathrm{rad}$), the $(4, 4)$ multipole can be safely neglected without significant loss of accuracy.

Our findings carry important implications for current and future \gls{GW} observations. For instance, they provide essential guidance for analysing possible light \gls{IMBH} candidates, including GW190521~\citep{LIGOScientific:2020iuh} and GW231123~\citep{LIGOScientific:2025rsn}. The systematic biases identified in our study could influence our ability to draw key astrophysical conclusions regarding their formation and whether or not they lie within the pair-instability mass gap~\citep{Woosley:2021xba}.

Although our work focuses specifically on light \gls{IMBH} systems, our recommendations can be broadly extrapolated for stellar-mass binary black hole systems. For example, our finding that the $(3, 3)$ multipole becomes important at \gls{SNR} $\gtrsim 20$ broadly aligns with its detection in GW190412 at \gls{SNR} $\approx 19$~\citep{LIGOScientific:2020stg}. As \gls{IMBH} sources represent a challenging case for the missing multipole problem, our results likely provide conservative lower bounds on the importance of starting frequency choice for lower-mass systems.

Several aspects of our study warrant further investigation. First, our analysis was limited by the {\textsc{NRSur7dq4}} model's calibrated parameter space and the current sensitivity of ground-based \gls{GW} detectors.
Extending this work to more extreme mass ratios and higher total masses would provide valuable guidance for next-generation detector science, which will naturally probe higher total masses due to their improved low-frequency sensitivity~\citep{Reitze:2019iox, Punturo:2010zz}. Second, our results are based on injections into zero-noise to isolate systematic effects. While this approach clearly demonstrates the missing multipole problem, real \gls{GW} observations contain detector noise that may partially mask or amplify these biases. Future studies should explore how realistic noise realizations interact with systematic errors arising from inappropriate starting frequencies. Finally, our analysis focuses on \gls{IMBH} systems, motivated by their enhanced higher-order multipole content. A natural follow-up study would investigate the missing multipole problem for stellar mass binaries. 

\section{Acknowledgements}

We thank Antoni Ramos-Buades for comments during the LIGO--Virgo--KAGRA internal review, Michael Williams for discussions during this project, and Xan Morice-Atkinson for answering \gls{HPC} questions. R.U and I.H acknowledge support from the Science and Technology Facilities Council (STFC) grant ST/V005715/1. C.H and L.K.N thank the UKRI Future Leaders Fellowship for support through the grant MR/T01881X/1, and C.H additionally thanks the University of Portsmouth for support through the Dennis Sciama Fellowship. We are grateful for computational resources provided by the SCIAMA \gls{HPC} cluster which is supported by the Institute of Cosmology and Gravitation (ICG) and the University of Portsmouth. We also thank the DiRAC Data Intensive service (DIaL) at the University of Leicester, managed by the University of Leicester Research Computing Service on behalf of the STFC DiRAC HPC Facility (www.dirac.ac.uk). The DiRAC service at Leicester was funded by BEIS, UKRI and STFC capital funding and STFC operations grants. DiRAC is part of the UKRI Digital Research Infrastructure.

This research has made use of data or software obtained from the Gravitational Wave Open Science Center (gwosc.org), a service of the LIGO Scientific Collaboration, the Virgo Collaboration, and KAGRA. This material is based upon work supported by NSF's LIGO Laboratory which is a major facility fully funded by the National Science Foundation, as well as the Science and Technology Facilities Council (STFC) of the United Kingdom, the Max-Planck-Society (MPS), and the State of Niedersachsen/Germany for support of the construction of Advanced LIGO and construction and operation of the GEO600 detector. Additional support for Advanced LIGO was provided by the Australian Research Council. Virgo is funded, through the European Gravitational Observatory (EGO), by the French Centre National de Recherche Scientifique (CNRS), the Italian Istituto Nazionale di Fisica Nucleare (INFN) and the Dutch Nikhef, with contributions by institutions from Belgium, Germany, Greece, Hungary, Ireland, Japan, Monaco, Poland, Portugal, Spain. KAGRA is supported by Ministry of Education, Culture, Sports, Science and Technology (MEXT), Japan Society for the Promotion of Science (JSPS) in Japan; National Research Foundation (NRF) and Ministry of Science and ICT (MSIT) in Korea; Academia Sinica (AS) and National Science and Technology Council (NSTC) in Taiwan. This material is based upon work supported by NSF's LIGO Laboratory which is a major facility fully funded by the National Science Foundation.

\emph{Software:} This work made use of \textsc{numpy}~\citep{harris2020array}, \textsc{scipy}~\citep{2020SciPy-NMeth} and \textsc{bilby}=2.2.2.1~\citep{Ashton:2018jfp,Romero-Shaw:2020owr}. Plots were prepared with \textsc{matplotlib}~\citep{2007CSE.....9...90H}, and {\textsc{pesummary}}~\citep{Hoy:2020vys}.

\emph{Author contribution}: R.U led the study and performed all Bayesian analyses in zero-noise. C.H conceived the project, supervised R.U and performed the analysis of GW231123\_135430. R.U and C.H wrote the paper. I.H proposed using the Mahalanobis Recovery Score to assess biases in Bayesian posteriors. I.H and L.K.N provided valuable comments on the manuscript and discussions throughout the project.

\section{Data Availability}
All posterior distributions are made available via \href{https://github.com/icg-gravwaves/missing_multipole_problem}{GitHub}. We also provide notebooks to reproduce the figures and code to calculate the Mahalanobis Recovery Scores.

\bibliographystyle{mnras}
\bibliography{refs}

\appendix

\section{Binary black hole notation}
\label{sec:appendix_notation}

Assuming a quasicircular orbit of two black holes with component masses $m_{1}$ and $m_{2}$, the total mass of the binary is $M = m_{1} + m_{2}$ and the mass ratio is $q = m_{2} / m_{1} \leq 1$. In some cases, the large mass ratio $Q = 1 / q \geq 1$ is used. Throughout this paper, we quote all masses and frequencies in the detector-frame. Source-frame masses, $m_{\mathrm{source}}$, are related to their detector frame quantities, $m_{\mathrm{det}}$, via $m_{\mathrm{source}} = m_{\mathrm{det}} / (1 + z)$ where $z$ is the cosmological redshift of the source.

The binary has orbital angular momentum $\mathbf{L}$, and spin angular momenta $\mathbf{S}$ such that $\mathbf{S} = \mathbf{S}_{1} + \mathbf{S}_{2}$. The magnitude of each spin vector is defined as $|\mathbf{S}_{i}| = m_{i}\chi_{i}$ where $\chi_{i}$ is bounded to be $\leq 1$, assuming the extremal Kerr limit $\chi_{i} = 1$~\citep{Kerr:1963ud, Kerr:2007dk}. The spin tilt denotes the angle between $\mathbf{S}_{i}$ and $\mathbf{L}$ for a given reference frequency, $\theta_{i} = |\mathbf{S}_{i} \cdot \mathbf{L}|$. 

Often it is convenient to describe the black hole spins by the effective inspiral~\citep{Ajith:2011ec} and precessing spin~\citep{Schmidt:2014iyl}, $\chi_{\mathrm{eff}}$ and $\chi_{\mathrm{p}}$ respectively. The effective inspiral spin represents the mass-weighted projection of the individual black hole spins onto $\mathbf{L}$. The effective spin is defined as

\begin{equation}
    \label{equ:effective_spin}
    \chi_\mathrm{eff} = \frac{(m_1 \mathbf{S}_{1} + m_2 \mathbf{S}_{2}) \cdot |\mathbf{L}|}{M},
\end{equation}
and
ranges from $-1$ (maximally anti-aligned spins) to $+1$ (maximally aligned spins). Positive values of $\chi_\mathrm{eff}$ slow down the inspiral in a phenomenom known as orbital hang-up~\citep{Campanelli:2006uy}. A value of $\chi_\mathrm{eff} = 0$ indicates that the mass-weighted average of the aligned spin components cancels out, which occurs if the spins lie primarily in the orbital plane, have unequal magnitudes, point in opposite directions, or are exactly zero.

The effective precessing spin characterises the projection of the  spin angular momentum perpendicular to the orbital angular momentum. This leads to precession of the individual black hole spins as well as the orbital plane~\citep{Apostolatos:1994mx}. Although the magnitude of the in-plane spin components oscillate due to nutation of the orbital plane~\citep{Lousto:2014ida}, their oscillation around a mean value is typically small. As such the level of precession is typically quantified by averaging the relative in-plane spin orientation,

\begin{equation}
    \chi_{\mathrm{p}} = \frac{1}{A_{1}}\max(A_{1}\mathrm{S}_{1\perp}, A_{2}\mathrm{S}_{2\perp}),
\end{equation}
where $A_{1}=2 + 3/2q$, $A_{2}=2+3q/2$, and $\mathrm{S}_{1\perp}$ and $\mathrm{S}_{2\perp}$ are the spin components perpendicular to the orbital angular momentum. $\chi_{\mathrm{p}}$ ranges between $0$ (zero-precession) and $+1$ (maximal precession). Other metrics have also been introduced to quantify the level of precession in a binary~\citep{Fairhurst:2019srr,Fairhurst:2019vut,Gerosa:2020aiw,Thomas:2020uqj}

The inclination angle, $\theta_{\mathrm{JN}}$, is defined as the angle between the line of sight to the observer and the total angular momentum vector of the binary system: $\mathbf{J} = \mathbf{L} + \mathbf{S}$.
A binary is said to be observed `face-on' when $\theta_{\mathrm{JN}}= 0$, or `edge-on' when $\theta_{\mathrm{JN}} = \pi / 2$.

A binary black hole will inspiral and eventually merge to form a perturbed Kerr black hole that continues to radiate \glspl{GW} through a superposition of exponentially damped \glspl{QNM}. The \gls{QNM} frequencies and damping times are functions of only the mass and spin of the unperturbed final Kerr black hole~\citep[see][for reviews]{Kokkotas:1999bd, Berti:2009kk, Franchini:2023eda}.
The frequency at which the two black holes merge is difficult to define. As such, the \gls{ISCO} frequency is often used to separate the inspiral regime from the merger and ringdown part of the signal~\citep[see e.g.][]{Ghosh:2016qgn,LIGOScientific:2021sio}. The \gls{GW} frequency corresponding to the \gls{ISCO} in Schwarzschild spacetime is,

\begin{equation}
    \label{eq:Sch_ISCO_freq}
    f_\mathrm{ISCO}^\mathrm{Sch}=  6^{-3/2}\left(GM\pi/c^3\right)^{-1}.
\end{equation}
The \gls{GW} frequency for an equatorial, prograde timelike orbit around a Kerr black hole of mass $M$ and
spin $\chi$ is a more accurate estimate for the \gls{ISCO} frequency of the binary~\citep{Ghosh:2016qgn},

\begin{equation}
    f_\mathrm{ISCO}^\mathrm{Kerr}= \frac{c^3 \sqrt{M}}{\pi G ({r_\mathrm{ISCO}^{3/2}+\chi\sqrt{M}})} \label{eq:Kerr_ISCO_freq},
\end{equation}
where $r_\mathrm{ISCO}$ is obtained by solving~\citep{Bardeen:1972fi}
\begin{equation}
    r(r-6M)+8\chi\sqrt{Mr}-3\chi^2 = 0 .\label{eq:ISCO_cubic}
\end{equation}

\section{Higher order multipole moments}
\label{sec:appendix_hms}

A \gls{GW} signal can be written as the complex sum of two polarizations, $h_{+}$ and $h_{\times}$: $h = h_{+} - ih_{\times}$. The -2 spin-weighted spherical harmonic decomposition can be used to express a \gls{GW} signal as an infinite sum of harmonics~\citep{Goldberg:1966uu,Thorne:1980ru}:
\begin{equation} \label{eq:harmonics}
    h_{+} - ih_{\times} = \sum_{\ell} \sum_{m = -\ell}^{\ell} {}^{-2}Y_{\ell m}h_{\ell m}.
\end{equation}
In this form, each multipole corresponds to a distinct angular structure in the radiation pattern and carries a portion of the \glspl{GW} total power.
The quadrupole $(\ell = 2)$
is the lowest-order contribution. For sources with spins aligned with the orbital angular momentum, Eq.~\ref{eq:harmonics} can be simplified as $h_{\ell m} = (-1)^{\ell} h_{\ell\, -m}^{*}$~\citep{Arun:2008kb,Ramos-Buades:2020noq}.

The amplitude of each higher-order multipole moment varies across the parameter space~\citep{Mills:2020thr}. Typically, the quadrupole $(\ell, m) = (2, 2)$ dominates the total power of the signal,
but this is not always the case: although the $(\ell, m) = (3, 3)$ and $(4, 4)$ multipoles have intrinsically lower amplitude than the $(2, 2)$, they extend to higher frequencies\footnote{The $(3, 3)$ and $(4, 4)$ multipoles will extend to frequencies that are $\sim 1.5$ and $2\times$ the merger frequency of the $(\ell, m) = (2, 2)$ respectively.}. As such, higher order multipoles can contribute significantly to the total power of the signal, and in some cases, dominate over the $(\ell, m) = (2, 2)$~\citep{Fairhurst:2023beb}.

The amplitude of each higher-order multipole primary depends on the total mass of the system, the mass ratio and the inclination angle of the binary~\citep{Mishra:2016whh,Mills:2020thr}. The importance of each higher order multipole relative to the quadrupole typically increases for more asymmetric binaries with higher total mass. Over most of the binary black hole parameter space, the $(3, 3)$ multipole is the most significant. However, the amplitude of the $(4, 4)$ increases rapidly as the total mass of the binary increases: for binaries with total mass above $\sim 75\, M_{\odot}$ and mass ratios $q > 0.5$, the $(4, 4)$ multipole is more significant than the $(3, 3)$~\citep{Mills:2020thr}. The dependence on the inclination angle varies for each higher order multipole. For instance, the $(2, 1), (3, 3)$ and $(4, 4)$ multipoles vanish for a binary observed face-on and are maximal for a binary observed edge-on.

\section{Mahalanobis Recovery Score and Difference} \label{sec:appendix_bias}

The Mahalanobis Recovery Score is a non-parametric method for quantifying biases in Bayesian posteriors by comparing different posterior realisations to the known value. By calculating the distance between the true value and the mean of different posterior realisations, a recovery score can be calculated. This method is similar to bootstrapping~\citep{Efron:1979bxm}.

Different posterior realisations for the probability distribution $P(\theta | d, m)$ can be produced by drawing $\mathcal{N}$ samples from the posterior, and placing a kernel at each data point. The mean of each kernel $\mu_{i}$ can be compared to the true value $\theta_{\mathrm{inj}}$ through the N-dimensional Mahalanobis distance~\citep{mahalanobis1936} $D_{i}(\theta_{\mathrm{inj}})$,
\begin{equation}
    \label{equ: Mahalanobis_distance}
    D_{i}(\theta_{\mathrm{inj}}) = \sqrt{(\theta_{\mathrm{inj}} - \mu_{i})^{\top} \Sigma^{-1} (\theta_{\mathrm{inj}} - \mu_{i})}
\end{equation}
where $(\theta_{\mathrm{inj}} - \mu_{i})$ is the vector from the true value, $\theta_{\mathrm{inj}}$, to the mean of each kernel, and $\Sigma^{-1}$ is the inverse covariance matrix of the original posterior~\citep{2015arXiv150402995F}.
If the Mahalanobis distance is less than a specified threshold, the mean is considered consistent with the known value. This is repeated for all posterior realisations, and the total number, $\mathfrak{N}$, where the Mahalanobis distance is less than the specified threshold is recorded. A recovery score can then be computed by dividing the $\mathfrak{N}$ by the total number of posterior realisations considered,
\begin{equation}
    r_{M}(\theta_{\mathrm{inj}}) = \mathfrak{N} / \mathcal{N}.
\end{equation}

The Mahalanobis recovery score is defined as the fraction of posterior realisations whose Mahalanobis distance from the injected value lies within a specified threshold corresponding to a chosen credible interval. For example, in this work we consider a threshold that corresponds to a $90\%$ credible interval for an N-dimensional multi-variate Gaussian distribution. The Mahalanobis recovery score is bounded between $0$ (no posterior realisations capture the truth) and a maximum that depends on the chosen threshold. In an ideal Gaussian case, the bound equals the chosen coverage probability (e.g. $0.9$ for a $90\%$ interval). A higher Mahalanobis recovery score therefore indicates better agreement between the posterior and the injected value, while a lower score highlights systematic bias. In our testing we found consistent trends between the Mahalanobis recovery score and the metric introduced in~\citep{Knee:2021noc}.

In order to compare analyses and/or different projections of the parameter space, we can compare the Mahalanobis recovery scores through the normalised

\begin{equation}
    \label{equ:heatmap_mahalanobis_score}
    \Delta r_{M} = \frac{r_M(\theta_{\mathrm{inj}}) - r'_M(\theta_{\mathrm{inj}})}{r_M(\theta_{\mathrm{inj}})},
\end{equation}
where $\Delta r_{M}$ is the Mahalanobis difference, $r_M(\theta_{\mathrm{inj}})$ is the recovery score for analysis 1 and $r'_M(\theta_{\mathrm{inj}})$ is the recovery score for analysis 2. When calculating the Mahalanobis difference,
we assume that $r_M(\theta_{\mathrm{inj}}) \geq r'_M(\theta_{\mathrm{inj}})$ so $\Delta r_{M} > 0$. There are some rare cases where our assumption is not valid, and Mahalanobis difference's between $-1 < \Delta r_{M} < 0$ are possible.

While $\Delta r_{M}$ allows us to quantitatively compare analyses 
it fails to tell us which analysis more accurately recovers the injected value. It nevertheless allows us to quantify trends and/or correlations between analyses: when $\Delta r_{M} = 0$, the posterior distributions for both analyses are comparable. When $\Delta r_{M} > 0$, analysis 1 generally performs better than analysis 2,
and in the limit of $\Delta r_{M} \rightarrow 1$, analysis 1 performs optimally, while analysis 2 consistently fails to recover the injected value.

We note that while the Mahalanobis recovery score accounts for the distribution’s covariance structure, it works best for distributions that are approximately Gaussian.

\section{Toy case} \label{sec:toy}

\begin{figure}
    \includegraphics[width=0.45\textwidth]{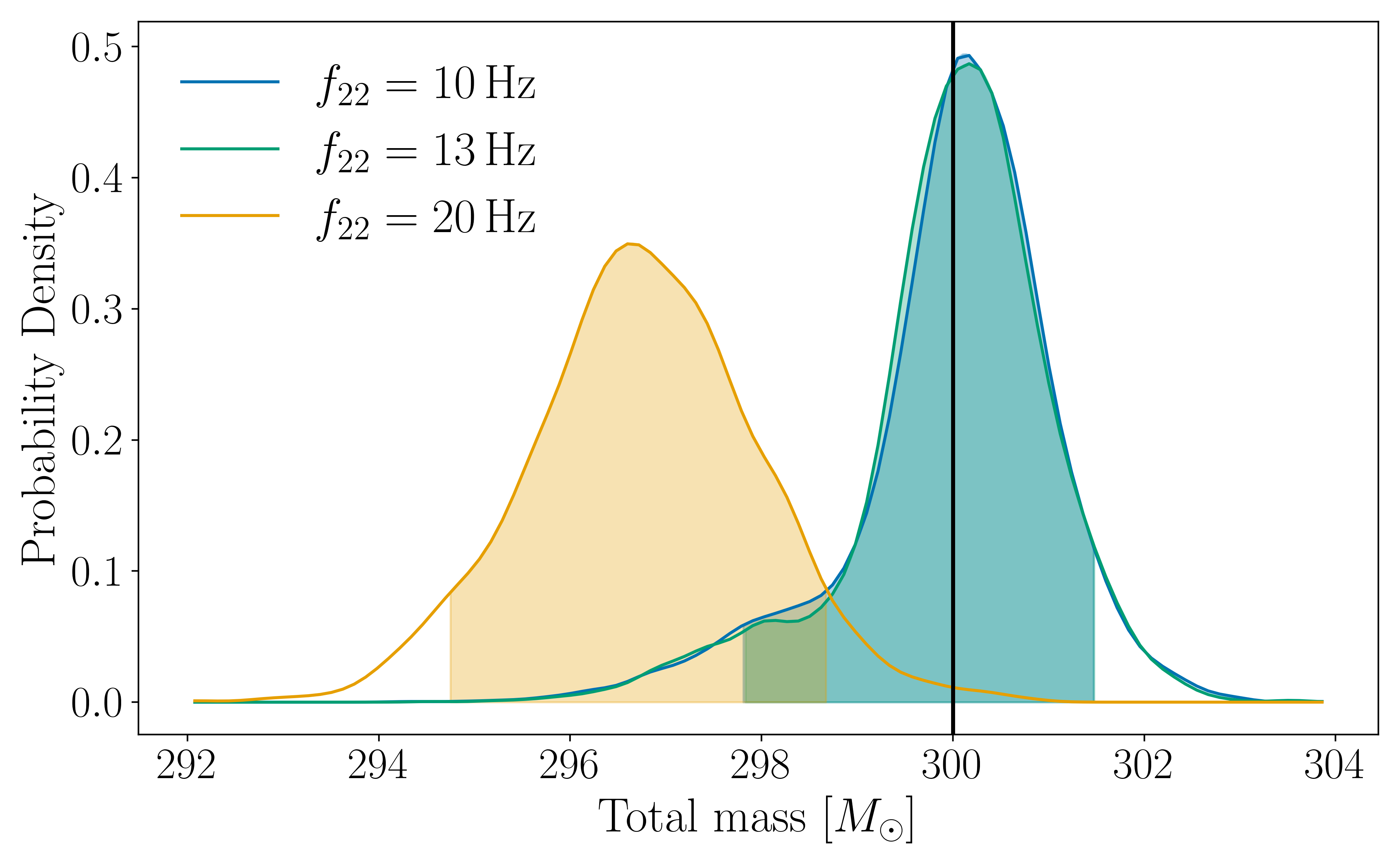}
    \caption{Comparison plot showing the inferred total mass posterior for a simulated \gls{GW} signal in zero-noise. We consider a simple two-dimensional problem where the model is characterised by only the binary component masses. The simulated signal, as well as all template waveforms, were generated with the {\textsc{NRSur7dq4}} waveform model~\citep{Varma:2019csw}. In blue, green, and orange we show the posterior distribution obtained when the initial frequency of the dominant quadrupole for the template is $f_{22} = 10, \ 13$, and $20\, \mathrm{Hz}$ respectively. The simulated signal had $f_{22} = 10\, \mathrm{Hz}$. In black we show the true value and the shaded region shows the 90\% credible interval.}
    \label{fig:toy}
\end{figure}

\setcounter{section}{5}
\begin{table*}
    \centering
    \begin{tabular}{|c|c|c|c|c|c|c|c|c|}
        \toprule
        \multirow{2}{*}{Parameter pair} & \multirow{2}{*}{$f_{22}\, [\mathrm{Hz}]$} & \multirow{2}{*}{Spin configuration} & \multicolumn{6}{c|}{$M\, [M_{\odot}]$}  \\
        & & & 200 & 250 & 300 & 350 & 400 & 450 \\
        \toprule
        \multirow{4}{*}{$\chi_1$ vs $\chi_2$} & \multirow{4}{*}{20} & Standard & 0.81 & 0.41 & 0.10 & 0.24 & 0.04 & 0.11 \\
        & & High spin magnitude & 0.72 & 0.79 & 0.82 & 0.84 & 0.88 & 0.76 \\
        & & Aligned-spin & 0.75 & 0.45 & 0.26 & 0.82 & 0.82 & 0.08 \\
        & & Highly precessing & 0.70 & 0.56 & 0.00 & 0.00 & 0.00 & 0.00 \\
        \midrule
        \multirow{4}{*}{$M$ vs $Q$} & \multirow{4}{*}{20} & Standard & 0.89 & 0.79 & 0.11 & 0.04 & 0.01 & 0.00 \\
        & & High spin magnitude & 0.84 & 0.87 & 0.71 & 0.06 & 0.41 & 0.00 \\
        & & Aligned-spin & 0.81 & 0.54 & 0.16 & 0.22 & 0.35 & 0.01 \\
        & & Highly precessing & 0.67 & 0.24 & 0.00 & 0.00 & 0.00 & 0.00 \\
        \bottomrule
    \end{tabular}
    \caption{The Mahalanobis recovery scores for the standard total mass series, see Sec.~\ref{sec:total_mass_series}, and the standard total mass series with increased spin magnitude, spins preferentially aligned with the orbital angular momentum and highly precessing configurations, see Sec.~\ref{subsubsec:high_spin_tilt}. We only show results for the $f_{22} = 20\, \mathrm{Hz}$ analyses, but scores for the other cases can be obtained in our public data release.}
    \label{tab:high_spin_tilt_mahal_scores}
\end{table*}

\begin{table*}
    \centering
    \begin{tabular}{|c|c|c|c|c|c|}
        \toprule
        \multirow{2}{*}{Parameter pair} & \multirow{2}{*}{$f_{22}\, [\mathrm{Hz}]$} & \multicolumn{4}{c|}{$\theta_{\mathrm{JN}}\, [\mathrm{rad}]$}  \\
        & & 0 & $\pi/6$ & $\pi/3$ & $\pi/2$ \\
         \toprule
        \multirow{3}{*}{$\chi_1$ vs $\chi_2$} & 10 & 0.69 & 0.74 & 0.82 & 0.81 \\
        & 13 & 0.69 & 0.67 & 0.60 & 0.71 \\
        & 20 & 0.04 & 0.05 & 0.10 & 0.60 \\
        \midrule
        \multirow{3}{*}{$M$ vs $Q$} & 10 & 0.85 & 0.86 & 0.89 & 0.88 \\
        & 13 & 0.89 & 0.69 & 0.52 & 0.81 \\
        & 20 & 0.00 & 0.04 & 0.11 & 0.30 \\
        \bottomrule
    \end{tabular}
    \caption{The Mahalanobis recovery scores for the inclination angle series, see Sec.~\ref{subsubsec:inclination_angle}.}
    \label{tab:inclination_angle_mahal_scores}
\end{table*}
\setcounter{section}{4}

Here, we consider a simple two-dimensional example, where the only parameters in the template are the binary component masses; all other parameters remain fixed.
We inject a \gls{GW} signal produced from a binary with total mass $M=300\, M_{\odot}$, mass ratio $q=0.9$ and inclination angle, defined as the angle between the line of sight and total angular momentum, $\theta_{\mathrm{JN}} = 1.22\, \mathrm{rad}$. The spin magnitudes for each black hole were $0.7$ and the spin tilt angles were $0.2$ and $0.8$ respectively. The luminosity distance was chosen such that the network \gls{SNR} was $\rho=75$. All other parameters were randomly chosen.

Fig.~\ref{fig:toy} shows the inferred total mass of the binary. As expected, when the template starts at $f_{22} = 10\, \mathrm{Hz}$ the injected value is well recovered.
We see that when the template starts at $f_{22} = 20\, \mathrm{Hz}$ missing power from \emph{e.g.} the $(\ell, m) = (3, 3)$ multipole between $20 - 30\,\mathrm{Hz}$ and the $(4, 4)$ multipole between $20 - 40\,\mathrm{Hz}$ causes the injected value to lie outside the 90\% credible interval. Given that a template with starting frequency $f_{22} = 13\, \mathrm{Hz}$ obtains comparable results to the $f_{22} = 10\, \mathrm{Hz}$ case, this indicates that missing power from the $(\ell, m) = (3, 3)$ multipole between $20 - 30\, \mathrm{Hz}$ causes significant biases in the inferred total mass of the binary. Although only a simple example, this clearly illustrates that care must be taken when analysing binary black holes at large \glspl{SNR}.

\section{Mahalanobis recovery score results} \label{sec:appendix_mahal_recov_tables}

The Mahalanobis recovery scores for the total mass series described in Secs.~\ref{sec:total_mass_series} and \ref{subsubsec:high_spin_tilt} are shown in Table~\ref{tab:high_spin_tilt_mahal_scores}, and the scores for the inclination angle run described in Section~\ref{subsubsec:inclination_angle} are shown in Table~\ref{tab:inclination_angle_mahal_scores}.

\bsp 
\label{lastpage}
\end{document}